\documentclass[useAMS,usenatbib]{mn2e}

\usepackage{color}
\usepackage{graphicx}
\usepackage{aas_macros}
\usepackage{times}
\usepackage{amsmath,amsfonts,amssymb}

\def\lesssim{\mathrel{\hbox{\rlap{\hbox{\lower4pt\hbox{$\sim$}}}\hbox{$<$}}}}
\def\gtrsim{\mathrel{\hbox{\rlap{\hbox{\lower4pt\hbox{$\sim$}}}\hbox{$>$}}}}

\def\unabla{{\bf \nabla}}
\def\uv{{\bf v}}

\def\ur{{\bf r}}

\def\wij{W_{ij}}

\def\dthr{\mathrm{d}^3{ r}}

\title[Entropy cores in non-radiative galaxy cluster simulations]
      {The formation of entropy cores in non-radiative galaxy cluster simulations: SPH versus AMR}
      \author[C. Power, J. I. Read \& A. Hobbs]{
        C. Power$^{1}$\thanks{chris.power@icrar.org}, J. I. Read$^{2}$ \& A. Hobbs$^{3}$\\
        $^1$International Centre for Radio Astronomy Research, University of Western Australia, 
        35 Stirling Highway, Crawley, Western Australia 6009, Australia\\
        $^2$Department of Physics, University of Surrey, Guildford, GU2 7XH, Surrey, United Kingdom\\
        $^3$Institute for Astronomy, Department of Physics, ETH Z\"urich, Wolfgang-Pauli-Strasse 16, 
        CH- 8093, Z\"urich, Switzerland}
\begin{document}

\date{}

\pagerange{\pageref{firstpage}--\pageref{lastpage}} \pubyear{2013}

\maketitle

\label{firstpage}

\begin{abstract} 
We simulate the formation and evolution of a massive galaxy cluster in a $\Lambda$CDM 
Universe using three different approaches to solving the equations of hydrodynamics in the 
absence of radiative cooling: one based on the `classic' Smoothed Particle Hydrodynamics 
(SPH) method; one based on a novel SPH algorithm with a higher order dissipation switch 
(SPHS); and one based on an adaptive mesh refinement (AMR) method. We find that SPHS and 
the AMR code are in excellent agreement with one another: in both, the spherically 
averaged entropy profile forms a well-defined core that rapidly converges with increasing 
mass and force resolution. By contrast, in agreement with previous work, SPH exhibits 
rather different behaviour. At low redshift, the entropy profile shows a systematic 
decrease with decreasing cluster-centric radius, converging on ever lower central entropy 
with increasing resolution. At higher redshift ($z \sim 1$), SPH is in better agreement 
with the other codes but shows much poorer numerical convergence. We trace the reason 
for these discrepancies to a known artificial surface tension in SPH that appears at phase 
boundaries. At early times, the passage of massive substructures close to the cluster centre
during its violent assembly stirs and shocks the gas to build up an entropy core. At late 
times, the artificial surface tension causes low entropy gas -- that ought to mix with the 
higher entropy gas -- to sink artificially to the centre of the cluster.

We use SPHS -- in which we can fully control the amount of numerical dissipation -- to 
study the contribution of numerical versus physical dissipation on the resultant entropy 
core. We argue that numerical dissipation is required to ensure single-valued fluid 
quantities in converging flows. However, provided this dissipation occurs only at the 
resolution limit, and provided that it does not propagate errors to larger scales, its 
effect is benign. There is no requirement to build `sub-grid' models of unresolved 
turbulence for galaxy cluster simulations. We conclude that entropy cores in non-radiative 
simulations of galaxy clusters are physical, resulting from entropy generation in shocked 
gas during the cluster assembly process. This finally puts to rest the long-standing 
puzzle of cluster entropy cores in AMR simulations versus their apparent absence in 
classic SPH simulations.

\end{abstract}


\section{Introduction}
\label{sec:intro}

Cosmological simulations are an established and powerful tool for 
studying the origin of cosmic structure and the formation of galaxies 
\citep[e.g.][]{2006Natur.440.1137S}. The formation and evolution of 
cosmic structure is assumed to be driven by a collisionless dark matter 
component, which forms massive collapsed structures -- so-called haloes 
-- that provide the potential wells within which gas cools and condenses 
to form galaxies \citep{1978MNRAS.183..341W}. The clustering and 
dynamics of the dark matter component has been studied in exhaustive 
detail over the last three decades and the $N$-body technique can be 
considered mature \citep[see, for example, the recent review 
of][]{2011EPJP..126...55D}. By contrast, the behaviour of the gas 
component is less secure. In part, this reflects uncertainty about the 
physical processes that are important 
\citep{2000ApJ...545..728T,2008ASL.....1....7M,2012MNRAS.423.1726S}, but 
it also reflects uncertainty about the manner in which the Euler 
equations are solved.

In particular, \citet{2007MNRAS.380..963A} found that the two most popular methods for 
solving the Euler equations in the literature -- Smoothed Particle 
Hydrodynamics (SPH; 
\citealt{1977AJ.....82.1013L,GingoldMonaghan1977,Monaghan1992}), and 
Adaptive Mesh Refinement (AMR; 
\citealt{1984JCoPh..53..484B,1989JCoPh..82...64B,1997ASPC..123..363B,1998JCoPh.143..519K,2000ApJS..131..273F,Teyssier2002}) 
-- give very different dissolution rates for a cold dense blob of gas 
moving at supersonic speed through a hot medium: in `classic' 
SPH\footnote{We define this to be the form of SPH implemented in the 
{\tt Gadget-2} code, and similar \citep{2005MNRAS.364.1105S}.}, the 
blobs survive much longer than their AMR counterparts. 
\citet{2010MNRAS.405.1513R} showed that this owes to two different 
problems with classic SPH: a leading order error in the momentum 
equation \citep{1999Dilts,2002JCoPh.179..238I}; and an artificial 
surface tension at phase boundaries 
\citep{2001MNRAS.323..743R,Price2008,2008MNRAS.387..427W}.

Over the past few years, there has been a welcome proliferation of new 
SPH ``flavours" and Lagrangian hydrodynamic methods designed to address the above problems \citep{2009arXiv0901.4107S, 2010MNRAS.406.2289H, 
2010arXiv1006.4159G, 2011MNRAS.tmp..196A, 2011MNRAS.417..136M, 
2012MNRAS.422.3037R, 2013MNRAS.428.1968K, 2013ApJ...768...44S, 
2013MNRAS.428.2840H}. These give significantly improved results on 
hydrodynamical test problems that have known analytic solutions 
\citep[e.g.][]{2012MNRAS.422.3037R}. When applied to astrophysical 
problems like galaxy formation, the results can also be quite different from the 
classic SPH simulations reported in the literature to date \citep{2012MNRAS.424.2999S,2012arXiv1207.3814H}. This suggests 
that -- in addition to the problem of unresolved or `sub-grid'\footnote{We refer, as is common in the literature, to physics below the resolution limit of a simulation as being `sub-grid' -- even though in SPH there are no actual grid cells. Such physics must either be omitted or modelled phenomenologically, with advantages and disadvantages to both approaches.} physics \citep{2012MNRAS.423.1726S} -- the choice of hydrodynamic solver matters.

Despite the above progress, a much older tension between SPH 
and AMR codes has eluded a complete explanation. 
\citet{1999ApJ...525..554F} simulated the formation of a non-radiative massive galaxy 
cluster using 12 different codes, finding 
that the SPH codes and the AMR\footnote{In fact, only one code in the 
study utilised adaptive mesh refinement techniques 
\citep{1995CoPhC..89..149B}. However, as a result, this was the only 
Eulerian mesh code that was capable of resolving the entropy core.} code 
converged on very different solutions from one another. In particular, 
the differences were most stark in the radial entropy profile of the 
gas, defined as:

\begin{equation}
	\label{eq:entropy}
	S(R) =\log \left[T_{\rm gas}(R)/\rho_{\rm gas}(R)^{2/3}\right]
\end{equation} 
where $R$ is the spherical radius with respect to the cluster centre of 
mass; $T_{\rm gas}$ is the gas temperature; and $\rho_{\rm gas}$ is the 
gas density. The SPH simulations appeared to converge on an ever lower 
central entropy as the force and mass resolution were increased, while 
the AMR simulation appeared to converge on a central constant entropy 
core. These results have been confirmed by several studies since 
\citep{2005MNRAS.364..909V, 2005ApJS..160....1O, 2005MNRAS.364..753D,2008MNRAS.387..427W, 2009MNRAS.395..180M}. 
\citet{2008MNRAS.387..427W} simulated gas in a galaxy cluster using
the \citet{1999ApJ...525..554F} initial conditions using classic SPH with
and without diffusion and concluded that lack of diffusion and particularly 
mixing in classic SPH gives rise to the non-convergent behaviour seen in 
earlier studies \citep[see also][]{Price2008}. \citet{2009MNRAS.395..180M} 
studied mergers between idealised galaxy clusters and traced the discrepancy 
to the artificial surface tension and the associated lack of multiphase 
fluid mixing in classic SPH, in agreement with \citet{2008MNRAS.387..427W}.
\citet{2011arXiv1109.3468S} arrive at a similar 
conclusion by comparing classic SPH with a new moving mesh code, {\tt Arepo}. However, while it is likely 
that the classic SPH result is incorrect, this does not automatically 
imply that the AMR results are correct. \citet{2011MNRAS.410..461V} report  
significant variation in the entropy profile for the same AMR 
code ({\tt Enzo}) when run with different refinement criteria, force 
resolution, and choice of energy equation. The refinement criteria 
appears to be most critical: depending on whether they refine on density 
or additionally on velocity jumps, they can produce entropy cores that 
differ in magnitude by up to a factor of two. Furthermore, this 
difference remains even when the numerical resolution is 
increased\footnote{This may simply imply that some refinement criteria 
are better than others. When comparing with fixed-grid simulations, 
\citet{2009MNRAS.395..180M} find that the standard density-refinement 
AMR gives excellent agreement. However, this test was performed only for 
a simplified set-up using just one single cluster merger; convergence is 
more difficult to achieve for the full cosmological case where additional substructure 
is present in the initial conditions as the resolution is increased. To 
our knowledge, comparisons between high resolution fixed grid and AMR 
simulations of non-radiative cosmological galaxy clusters have not yet been 
performed}. In 
addition to variations in the entropy profile due to a 
particular flavour of AMR, differences are also seen when comparing the 
AMR results to that of the moving mesh code {\tt Arepo}. 
\citet{2009arXiv0901.4107S} report an entropy core that is significantly 
lower than that found in AMR codes (e.g. compare Figure 45 of 
\citealt{2009arXiv0901.4107S} with Figure 18 of 
\citealt{1999ApJ...525..554F} or Figure 5 of 
\citealt{2005MNRAS.364..909V}).

The above discrepancies between different numerical techniques are 
important. Since the advent of space-based X-ray satellites, it has been 
known that real galaxy clusters split into two broad observational 
classes: {\it cooling-core} (CC) clusters that have very low central 
entropy, and {\it non-cooling-core} (NCC) clusters that have an 
approximately constant entropy core in the centre 
\citep[e.g.][]{2008MNRAS.386.1309M}. If different methods for solving 
the non-radiative Euler equations lead to a CC (SPH) or NCC (AMR) cluster, then we are 
unable to determine the real physical processes that drive this 
dichotomy in nature. A proper understanding of the thermodynamic state 
of cluster gas is vital for using clusters as cosmological probes 
\citep[e.g.][]{1972ApJ...176....1G,2005RvMP...77..207V}; as probes of 
the baryon content of the Universe \citep[e.g.][]{2009ApJ...703..982G}; 
or as probes of dark matter through their hot X-ray emitting gas 
\citep[e.g.][]{1976A&A....49..137C,1989ApJ...337...21H,2006ApJ...640..691V}.

In this paper, we revisit the problem of modelling non-radiative  
cosmological galaxy clusters using a new flavour of SPH -- SPHS -- that 
is designed to resolve two key problems with SPH: (i) multivalued 
pressures at flow boundaries that lead to a numerical surface tension; 
and (ii) poor force accuracy in shearing flows 
\citep{2012MNRAS.422.3037R}. The former problem is cured by introducing 
a higher order dissipation switch that detects, in advance, when 
particles are going to converge\footnote{This is similar to a switch proposed 
first by \citet{2010MNRAS.408..669C} but switching on {\it all} advected fluid 
quantities, not just the artificial viscosity.}. If this happens, conservative 
dissipation is switched on for all advected fluid quantities (i.e. 
artificial thermal conductivity, artificial viscosity, etc.). The 
dissipation is switched off again once particles are no longer 
converging. This ensures that all fluid quantities are single-valued 
throughout the flow by construction. The second problem is cured by 
moving to higher order stable kernels that can support larger neighbour 
numbers \citep{2010MNRAS.405.1513R,2012arXiv1204.2471D}. We use the default kernel choice from 
\citet{2010MNRAS.405.1513R}: the HOCT4 kernel with 442 neighbours. \citet{2012MNRAS.422.3037R} 
demonstrated that SPHS performs very well on a broad range of 
hydrodynamic test problems including the Sod shock tube, Sedov-Taylor 
blast wave, Gresho vortex, and the high density contrast 
Kelvin-Helmholtz instability test, giving excellent agreement with 
analytic expectations. A key advantage of the SPHS method is that 
we can explicitly control the amount of numerical dissipation. This 
allows us to measure how dissipation at the resolution limit feeds back 
to larger resolved scales in the simulation. 

We test the convergence of our results with increasing mass and force resolution, the sensitivity 
to the numerical dissipation parameters, and present explicit 
comparisons with an AMR code {\tt RAMSES} \citep{Teyssier2002}. In 
performing these numerical experiments, we seek to address three key 
questions:
\begin{enumerate} 
\item What is the origin of the discrepancy between the classic 
SPH and the AMR results?
\item Do resolved scales in non-radiative 
simulations of galaxy cluster formation care about the details of 
dissipation (physical or numerical) on unresolved scales?
\item What is the role of gravitational shock heating as an entropy generation 
mechanism in galaxy clusters?
\end{enumerate}

This paper is organised as follows. In \S\ref{sec:methods}, we describe the 
different numerical methods -- `classic' SPH, SPHS and AMR -- used in this work. 
In \S\ref{sec:sims}, we describe our simulation suite. In \S\ref{sec:results}, we 
present our results. In \S\ref{sec:discussion}, we return to the three key questions 
posed above and discuss the meaning of our results for real galaxy clusters in the 
Universe. Finally, in \S\ref{sec:conclusions} we present our conclusions.

\section{Methods}
\label{sec:methods}

\subsection{`Classic' SPH}\label{sec:sph} 

We adopt the fully conservative `entropy' form of SPH described in 
\citet{2002MNRAS.333..649S}. The discretised Euler equations are: 
\begin{equation}
  \rho_i = \sum_j^N m_j \wij(|{\bf r}_{ij}|,h_i)
  \label{eq:continuity}
\end{equation}

\begin{equation}
\frac{d\uv_i}{dt} = -\sum_j^N m_j \left[f_i \frac{P_i}{\rho_i^2} \nabla_i W_{ij}(h_i) + f_j \frac{P_j}{\rho_j^2} \unabla_i W_{ij}(h_j)\right]
\label{eq:momentum}
\end{equation}

\noindent and:

\begin{equation}
  P_i = A_i \rho_i^\gamma
  \label{eq:equation_of_state}
\end{equation}

\noindent Here $m_i$ is the mass of particle $i$; ${\bf v}_i$ is the velocity; $P_i$ is the 
pressure; $\rho_i$ is the density; $A_i$ is a function that is monotonically related to the 
entropy (hereafter referred to as the `entropy'); $W$ is a symmetric kernel that obeys the 
normalisation condition:

\begin{equation}
  \int_{V} W(|\ur-\ur'|,h) \dthr' = 1 ,
  \label{eq:normw}
\end{equation}
and the property (for smoothing length $h$):
\begin{equation}
  \lim_{h\rightarrow 0} W(|\ur-\ur'|,h) = \delta(|\ur-\ur'|),
\end{equation}
where ${\bf r}_{ij} = {\bf r}_j - {\bf r}_i$ is the vector position of 
the particle relative to the centre of the kernel; and the function 
$f_i$ in equation \ref{eq:momentum} is a correction factor that ensures energy 
conservation for varying smoothing lengths:

\begin{equation}
  f_i = \left(1 + \frac{h_i}{3\rho_i}\frac{\partial \rho_i}{\partial h_i}\right)^{-1}
\end{equation}
Note that we do not use the above conservative momentum equation in SPHS 
since it leads to larger force errors with only a modest improvement in 
energy conservation (at least when applied to galaxy and galaxy cluster 
formation simulations; see \citet{2010MNRAS.405.1513R} and 
\citet{2012MNRAS.422.3037R} for further details).

We use a variable smoothing length $h_i$ as in 
\cite{2002MNRAS.333..649S} that is adjusted to obey the following 
constraint equation:

\begin{equation}
\frac{4\pi}{3} h_i^3 n_i = N_n \qquad ;\,\mathrm{with} \qquad n_i = \sum_j^N W_{ij}
\label{eq:fixedmass}
\end{equation}
where $N_n$ is the typical neighbour number (the number of particles 
inside the smoothing kernel, $W$). The above constraint equation gives 
fixed mass inside the kernel if particle masses are all equal. We use a 
standard cubic spline (CS) kernel with $N_n = 40$ neighbours.

There is no dissipation switching and $\alpha = \alpha_\mathrm{max} = 
\mathrm{const.} = 1$ always. There is also no dissipation in entropy; 
the only numerical dissipation applied is the artificial viscosity. This 
prevents multivalued momenta from occurring, but not multivalued entropy 
or pressure \citep[e.g.,][]{2010MNRAS.405.1513R}.

\subsection{SPH with a higher order dissipation Switch (SPHS)}\label{sec:SPHS} 

SPHS minimises force errors in the discretised hydrodynamic equations of 
motion \citep[cf.][]{2010MNRAS.405.1513R}, and so the key difference 
with respect to classic SPH arises in the momentum equation, which is 
recast as:

\begin{equation}
  \frac{d\uv_i}{dt} = -\sum_j^N \frac{m_j}{\rho_i\rho_j} \left[P_i + P_j\right] \nabla_i \overline{W}_{ij}
  \label{eq:sphs_momentum}
\end{equation}
where $\overline{W}_{ij} = \frac{1}{2}\left[W_{ij}(h_i) + W_{ij}(h_j)\right]$ is a symmetrised smoothing kernel.

We adopt the `HOCT4' smoothing kernel with 442 neighbours as this gives 
significantly improved force accuracy and convergence 
\citep{2010MNRAS.405.1513R,2012MNRAS.422.3037R}:

\begin{equation}
W = \frac{N}{h^3}\left\{\begin{array}{lr}
Px + Q & 0 < x \le \kappa \\
(1-x)^4 + (\alpha - x)^4 + (\beta-x)^4 & \kappa < x \le \beta \\
(1-x)^4 + (\alpha - x)^4 & \beta < x \le \alpha\\
(1-x)^4 & x \le 1 \\
0 & \mathrm{otherwise} \end{array}\right.
\label{eq:hoctkern}
\end{equation}
with $N = 6.515$, $P=-2.15$, $Q=0.981$, $\alpha = 0.75$, $\beta = 0.5$ and $\kappa = 0.214$.

In addition to the above equations of motion, numerical dissipation is 
switched on if particles are converging. This avoids multivalued fluid 
quantities occurring at the point of convergent flow. Without such 
dissipation, the resulting multivalued pressures drive waves through the 
fluid that propagate large numerical errors and spoil convergence. The 
switch is given by:

\begin{equation}
\alpha_{\mathrm{loc},i} = \left\{
\begin{array}{lr}
\frac{h_i^2 |\unabla(\unabla \cdot {\bf v}_i)|}{h_i^2 |\unabla(\unabla \cdot {\bf v}_i)| + h_i |\unabla \cdot {\bf v}_i|+ n_s c_i} \alpha_\mathrm{max} & \unabla \cdot {\bf v}_i < 0 \\
0 & \mathrm{otherwise} 
\end{array}\right.
\label{eq:alphalocv}
\end{equation}
where $\alpha_{\mathrm{loc},i}$ describes the amount of dissipation for 
a given particle in the range $[0,\alpha_\mathrm{max} = 1]$; $c_i$ is 
the sound speed of particle $i$; and $n_s = 0.05$ is a `noise' parameter 
that determines the magnitude of velocity fluctuations that trigger the 
switch. Equation \ref{eq:alphalocv} turns on dissipation if $\unabla 
\cdot {\bf v}_i < 0$ (convergent flow) and if the magnitude of the 
spatial derivative of $\unabla \cdot {\bf v}_i$ is large as compared to 
the local divergence (i.e., if the flow is going to converge). The key 
advantage as compared to most other switches in the literature is that 
it acts as an early warning system, switching on {\it before} large 
numerical errors propagate throughout the fluid \citep[see 
also][]{2010MNRAS.408..669C}. The second derivatives of the velocity 
field are calculated using high order polynomial gradient estimators 
described in \citet{2003ApJ...595..564M} and 
\citet{2012MNRAS.422.3037R}. We use the above switch to turn on 
dissipation in all advected fluid quantities -- i.e., the momentum 
(artificial viscosity) and entropy (artificial thermal conductivity). 
Once the trajectories are no longer converging, the dissipation 
parameter decays back to zero on a timescale $\sim h_i / c_i$. The 
dissipation equations are fully conservative and described in detail in 
\cite{2012MNRAS.422.3037R}. The key free numerical parameter is 
$\alpha_\mathrm{max}$ that sets the rate of dissipation that occurs 
when particle trajectories attempt to cross. This parameter allows us 
to control the amount of resolution-scale dissipation from zero (similar 
to classic SPH), up to large values ($\alpha_\mathrm{max} \sim 1$ is the 
natural choice since this leads to single-valued fluid quantities on a 
timescale comparable to the particle trajectory convergence time). We might 
hope that the results on resolved scales do not care about the amount or 
form of dissipation that occurs at the resolution limit, since such dissipation 
moves to ever smaller scales as the resolution is increased. Indeed, in test 
problems SPHS converges independently of the choice of $\alpha_\mathrm{max}$ so 
long as it is large enough to avoid multi-valued pressures 
\citep{2012MNRAS.422.3037R}. However, for complex non-linear problems this is 
not entirely clear. Dissipation on unresolved scales could in principle affect 
the results on resolved scales if it causes an upwards transfer of information 
in the form of pressure waves, for example. This would manifest as numerically 
`converged' results on resolved scales that depend on the magnitude and form of 
the numerical dissipation parameters. We test this explicitly using SPHS in \S\ref{sec:results}.

\subsection{AMR}\label{sec:amr}

We used the {\small RAMSES} adaptive mesh refinement (AMR) code of 
\citet{Teyssier2002}. The evolution of the gas is followed using a 
second-order unsplit Godunov scheme for the Euler equations. 
Collisionless $N$-body particles are evolved using a particle-mesh 
solver with a Cloud-In-Cell interpolation. The coarse mesh is refined 
using a quasi-Lagrangian strategy, such that cells are refined when more 
than 8 dark matter particles lie in a cell or if the baryon density is 
larger than 8 times the initial dark matter resolution. When refined, 
cells are divided into 8 smaller cubic cells, giving a factor of 2 
increase in spatial resolution. These smaller cells may be refined 
further, up to a maximum level of refinement defined by the user. 
Timesteps are adapted to the levels of refinement so that the timestep 
for cells at refinement level $\ell$ is twice as long as the timestep at 
level $\ell+1$.

\section{Simulations}
\label{sec:sims}

\paragraph*{Parent Simulation} Our parent $N$-body simulation follows 
structure formation in a periodic volume of side $L_{\rm box}=150 h^{-1} 
\rm Mpc$ containing $150^3$ particles in the $\Lambda$CDM model with 
cosmological $\Omega_0=0.7$, $\Omega_{\Lambda}=0.3$, $h=0.7$, and a 
normalisation of $\sigma_8=0.9$. The particle mass $m_p \simeq 8.3 
\times 10^{10} h^{-1} \rm M_{\odot}$ ensures that the most massive 
clusters likely to form in a volume of this size will contain $\sim 
10^4$ particles within their virial radius $r_{\rm vir}$ at $z$=0; this 
is sufficient to define the region to be resimulated at higher mass 
resolution.

We used the parallel TreePM code {\small GADGET2} 
\citep{2005MNRAS.364.1105S} with constant comoving gravitational 
softening $\epsilon=20 h^{-1} \rm kpc$ to run the simulation and 
constructed group catalogues using {\small 
AHF}\citep[{\small{\textbf{A}MIGA}'s
  \small{\textbf{H}}alo \small{\textbf{F}}inder}; cf.][]{2009ApJS..182..608K}.
For each halo in the AHF catalogue we determined the centre-of-density 
$\vec{r}_{\rm cen}$ using the iterative ``shrinking spheres'' method described 
in \citealt{power.etal.2003} and we identified this as the halo centre. From 
this, we calculated the halo's virial radius $r_{\rm vir}$, which we define 
as the radius within which the mean interior density is $\Delta_{\rm vir}$ 
times the critical density of the Universe at that redshift, 
$\rho_{\rm c}(z)=3H^2(z)/8\pi G$, where $H(z)$ and $G$ are the Hubble parameter 
at $z$ and the gravitational constant respectively. The corresponding virial 
mass $M_{\rm vir}$ is 

\begin{equation}
  \label{eq:mvir}
        {M_{\rm vir}=\frac{4\pi}{3} \Delta_{\rm vir} \rho_{\rm c} r_{\rm vir}^3.}
\end{equation}

\noindent where we adopt $\Delta_{\rm vir}$=200, independent of 
redshift.

\paragraph*{Galaxy Cluster Resimulation} We chose to resimulate the most 
massive halo to form in our parent simulation with both gas and dark 
matter -- corresponding to a galaxy cluster with a virial mass of 
$M_{\rm vir} \simeq 6 \times 10^{15} h^{-1} \rm M_{\odot}$ ($N_{\rm vir} 
\simeq 7200$ particles) and virial radius of $R_{\rm vir} \simeq 1.35 
h^{-1} \rm Mpc$ at $z$=0. The resimulation technique allows us to target 
our computational effort so that we can employ high mass and force 
resolution in a sub-volume of the original parent simulation, whilst 
also capturing the large scale tidal effects due to all the other matter 
in the Universe. To set up the initial conditions for our resimulations, 
we took the following steps;

\begin{enumerate}

\item We identified all particles within a volume of radius
  $\sim 3.5 R_{\rm vir}$ centred on the centre of density $\vec{r}_{\rm 
  cen}$ of the cluster halo at $z$=0 in the parent simulation and 
  determined their positions in its initial conditions at the starting 
  redshift $z_{\rm start}$=74.
  
\item Using the particle velocities and $z_{\rm start}$, we applied an 
  inverse Zel'dovich transformation to obtain the particle positions at 
  $z=\infty$, from which we determined the spatial extent of the initial 
  Lagrangian volume. This volume defines the central region of a 
  multi-level mask for the high resolution region.
  
\item We populated this simulation volume with particles with a number 
  density set by our high resolution mask; the number density of particles 
  within the central region of the mask is highest -- set by the desired 
  mass resolution of the resimulation -- and declines in subsequent levels 
  of the mask, such that the mass resolution coarsens with increasing 
  distance from the central region. For hydrodynamical simulations of the 
  kind described in this paper, we include both gas and dark matter 
  particles within the central region, with number densities fixed by the 
  cosmological baryon and dark matter density pararmeters $\Omega_{\rm b}=0.04$ 
  and $\Omega_{\rm DM}=0.26$.
  
\item We imposed two sets of density perturbations on this composite particle
  distribution. The first set correspond to the original set of perturbations 
  that were present in the initial conditions of the parent simulation, with
  minimum and maximum wavenumbers, $k_{\rm min}=2\pi/L_{\rm box}$ and
  $k_{\rm max}=\pi\,N_p/L_{\rm box}$, and the second set corresponds to 
  perturbations that were not present in the initial conditions, 
  $k_{\rm min}=2\pi/L_{\rm hires}$ and $k_{\rm max}=\pi\,N_{\rm hires}/L_{\rm hires}$.
  Here $L_{\rm box}$ and $L_{\rm hires}$ are the side-lengths of the parent 
  volume and box encompassing the high resolution patch respectively, and 
  $N_p$ and $N_{\rm hires}$ are the number of dark matter particles
  on a side in these boxes.
  
\item From these perturbations we constructed the initial baryon density
  perturbation field $\delta_b(\vec{x})=\rho/\bar{\rho}-1$ at mesh points 
  $\vec{x}$ and used the Zel'dovich approximation to compute the velocity 
  field $\vec{v}(\vec{x})$. From these we initialised gas and dark matter 
  particle positions and velocities.

\end{enumerate}

We used the TreePM $N$-body SPH code {\small GADGET3}, in which we have 
implemented SPHS, to run the bulk of our simulations. Gravitational force 
softenings for both the dark matter and the gas were chosen in accordance 
with the optimal criterion of $\epsilon_{\rm opt}=4\,r_{\rm vir}/\sqrt{N_{\rm vir}}$ 
of \citet{power.etal.2003}. 

For comparison, we also ran a subset of the simulations using the public 
version of the AMR code {\small RAMSES}. {\small RAMSES} takes as input
baryon density perturbation and velocity fields on uniform cubic meshes --
the coarser meshes capture the influence of the large-scale gravitational 
field while the finest mesh -- in combination with a refinement map that
tells {\small RAMSES} where to place its initial refinements -- corresponds to the 
high resolution region. We ran two simulations -- one with a minimum level of
refinement of $\ell$=7, the other with $\ell$=8, and in both cases we fixed the
maximum level of refinement at $\ell$=15. These correspond to mesh cell lengths
of $\Delta$=$L_{\rm hires}/2^{\ell}$$\sim$0.3 (0.16) $h^{-1} \rm Mpc$ for $\ell$=7 (8),
and $\Delta \sim 0.001 h^{-1} \rm Mpc$ for $\ell$=15. In both cases we used
a criterion of {\tt m\_refine}=8 (dark matter particles or factor increase
in the initial gas mass resolution) to trigger new refinements.

\section{Results}
\label{sec:results}

In the following subsections we compare the results of our SPH, SPHS and 
AMR runs. The focus of our analysis is on the entropy of the cluster 
gas, which we define according to Equation~(\ref{eq:entropy}). We 
construct our spherically averaged entropy profiles by defining the 
cluster centre of density $\vec{r}_{\rm cen}$ using the shrinking 
spheres method \citep[cf.][]{power.etal.2003}, sorting particles by 
cluster-centric radius, and assigning them to 25 spherical logarithmic 
bins equally spaced between $R_{\rm min}=0.01 R_{\rm vir}$ and $R_{\rm 
vir}$, where $R_{\rm vir}$ is defined according to 
Equation~(\ref{eq:mvir}).

\subsection{Comparison of SPH and SPHS}

%%% Visual inspection of the cluster at z=0 on large-scales (Fig 1) and intermediate-to-small
%%% scales (Fig 2) in SPH and SPHS.

\begin{figure*}
  \centerline{\includegraphics[width=8cm]{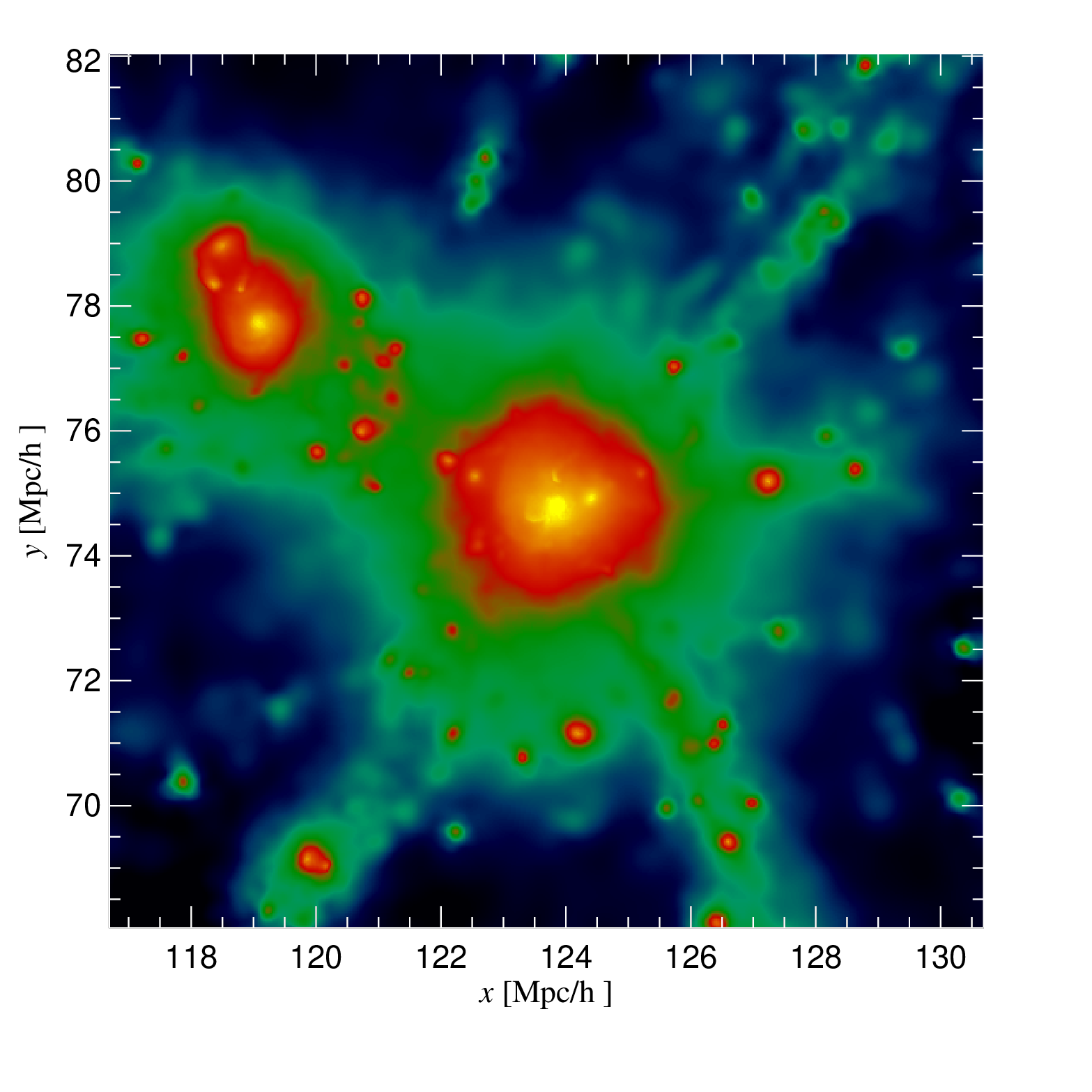}
    \includegraphics[width=8cm]{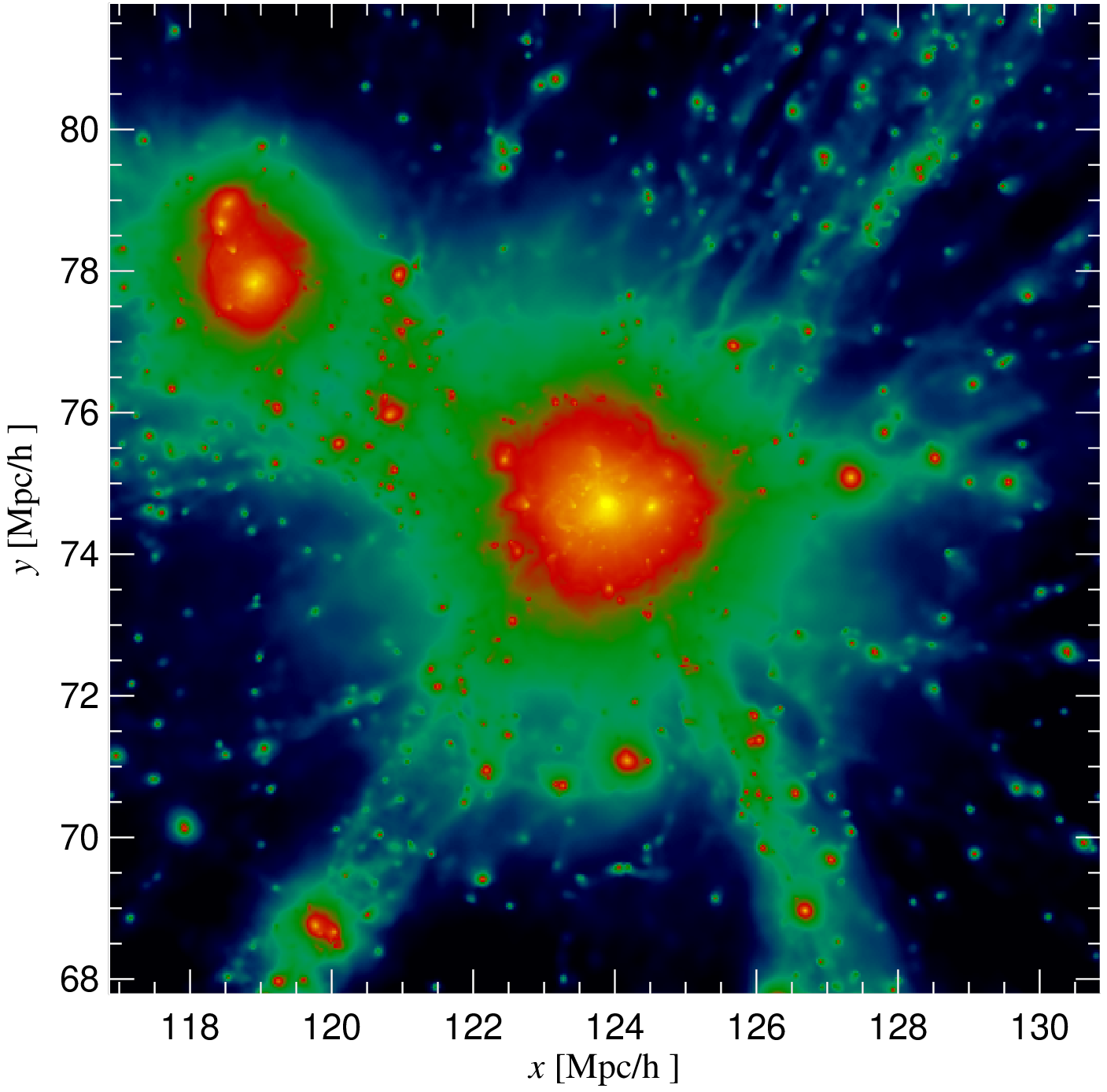}}
  \centerline{\includegraphics[width=8cm]{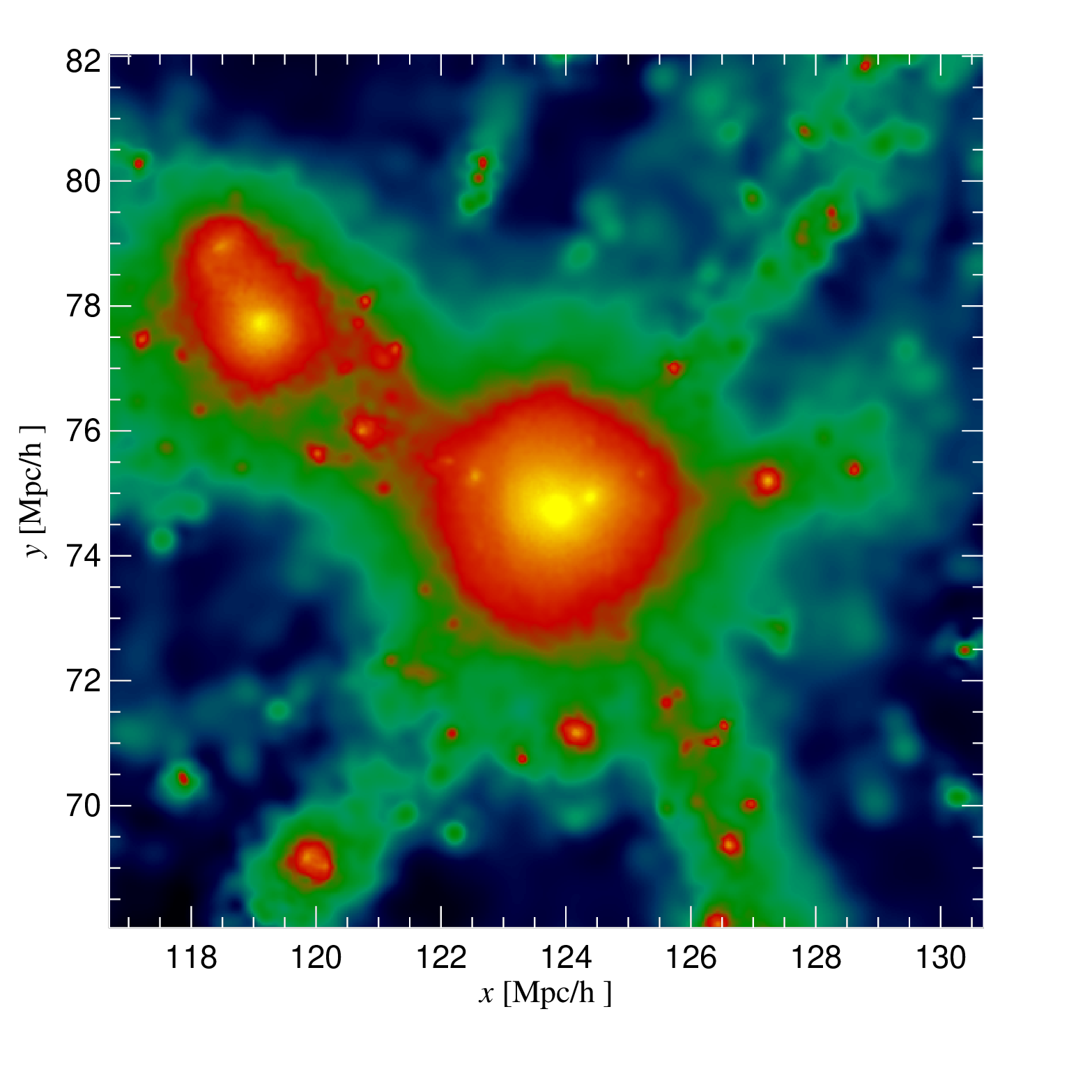}
    \includegraphics[width=8cm]{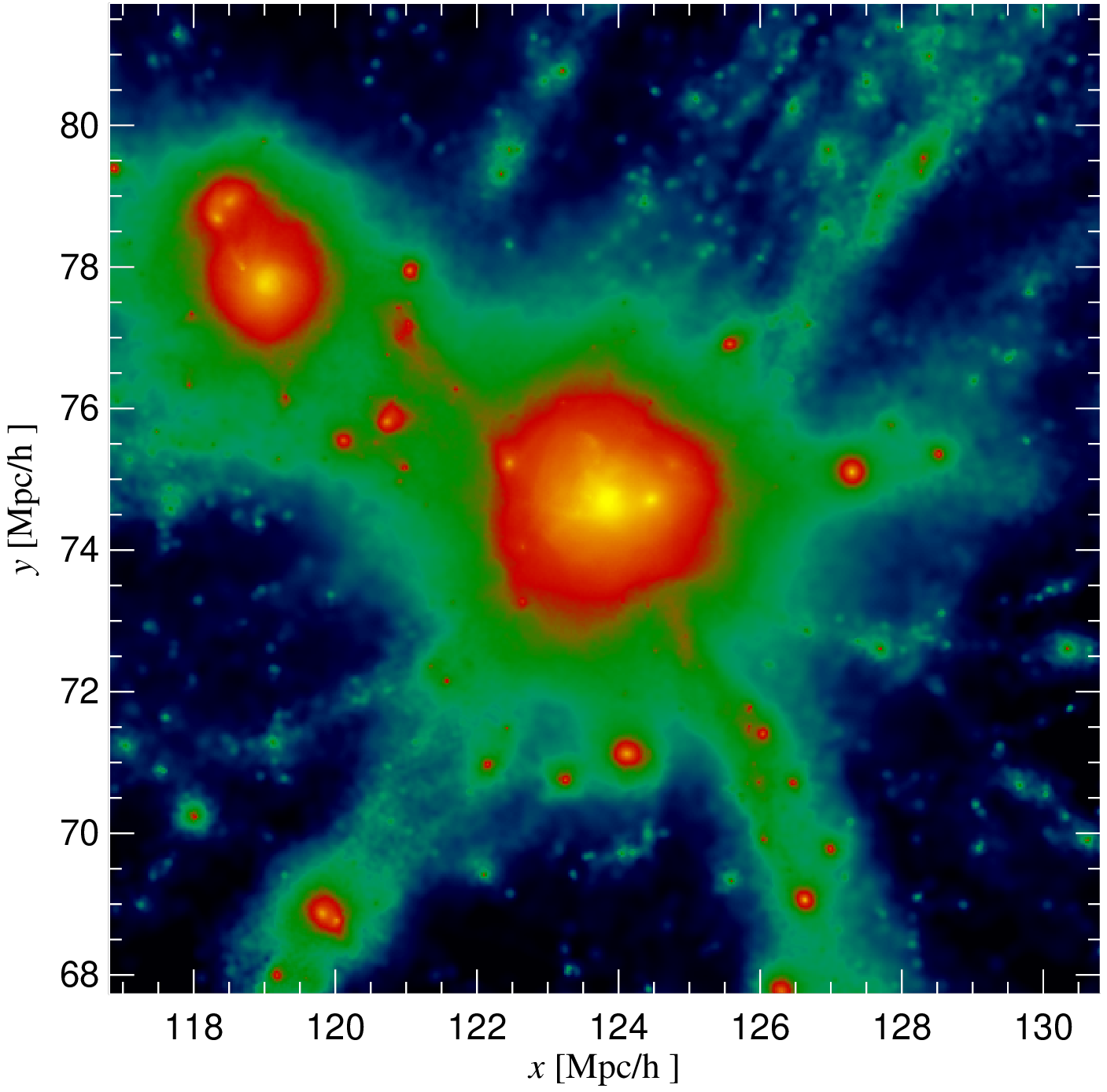}}
  \caption{\textbf{Visual Impression.} Projected gas density maps in a cube of 
    side 15 $h^{-1} {\rm Mpc}$ centred on the cluster at $z$=0 in the $\times$32 
    and $\times$128 SPH (top left and right) and SPHS (bottom left and right) runs 
    respectively.}
\label{fig:SPHS_vs_SPH}
\end{figure*}

\begin{figure*}
  \centerline{
    \includegraphics[width=8cm, clip, trim = 0cm 1.2cm 0cm 0cm]{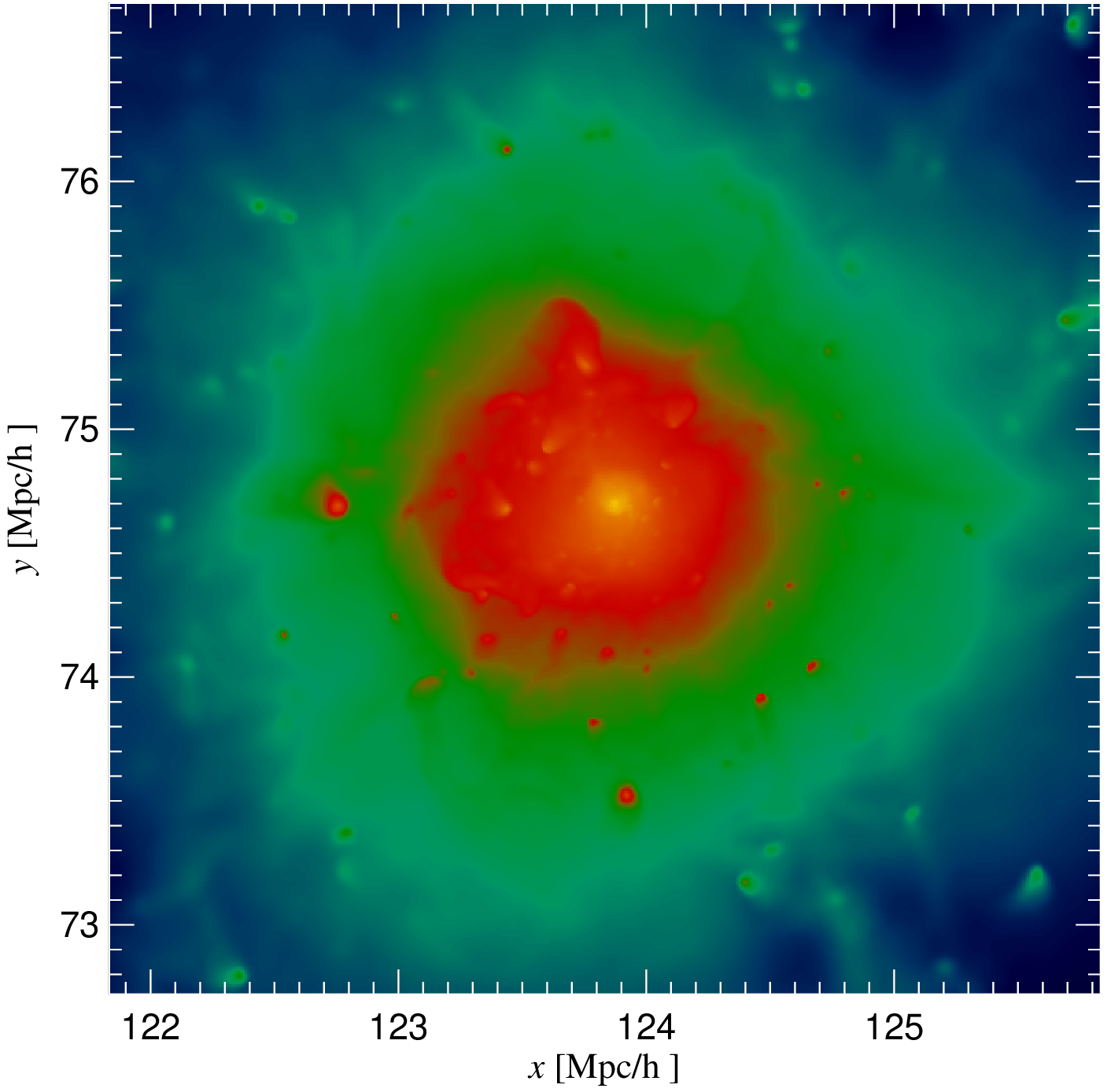}
    \includegraphics[width=8cm, clip, trim = 0cm 1.2cm 0cm 0cm]{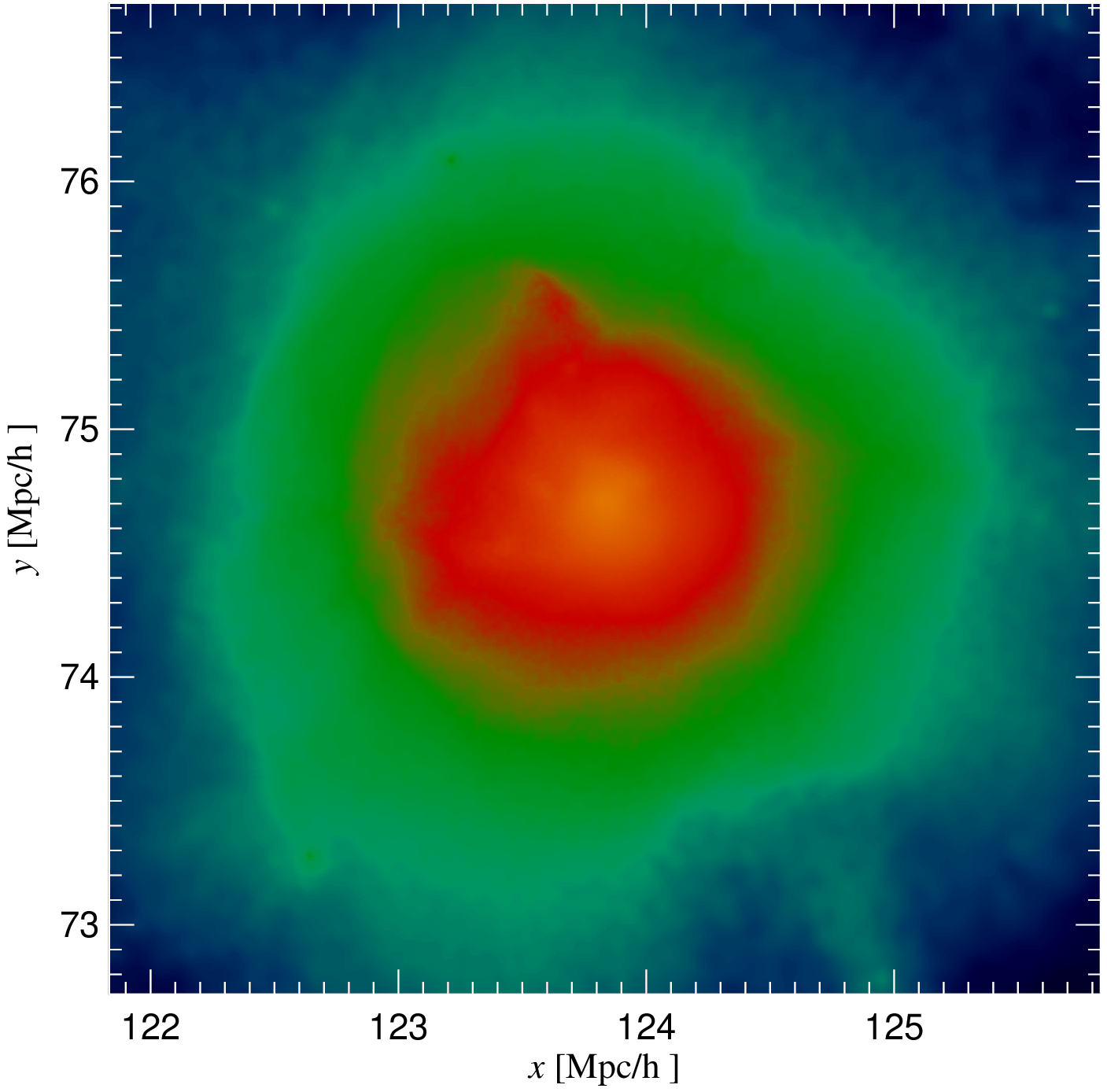}}
  \centerline{
    \includegraphics[width=8cm, clip, trim = 0cm 1.2cm 0cm 0cm]{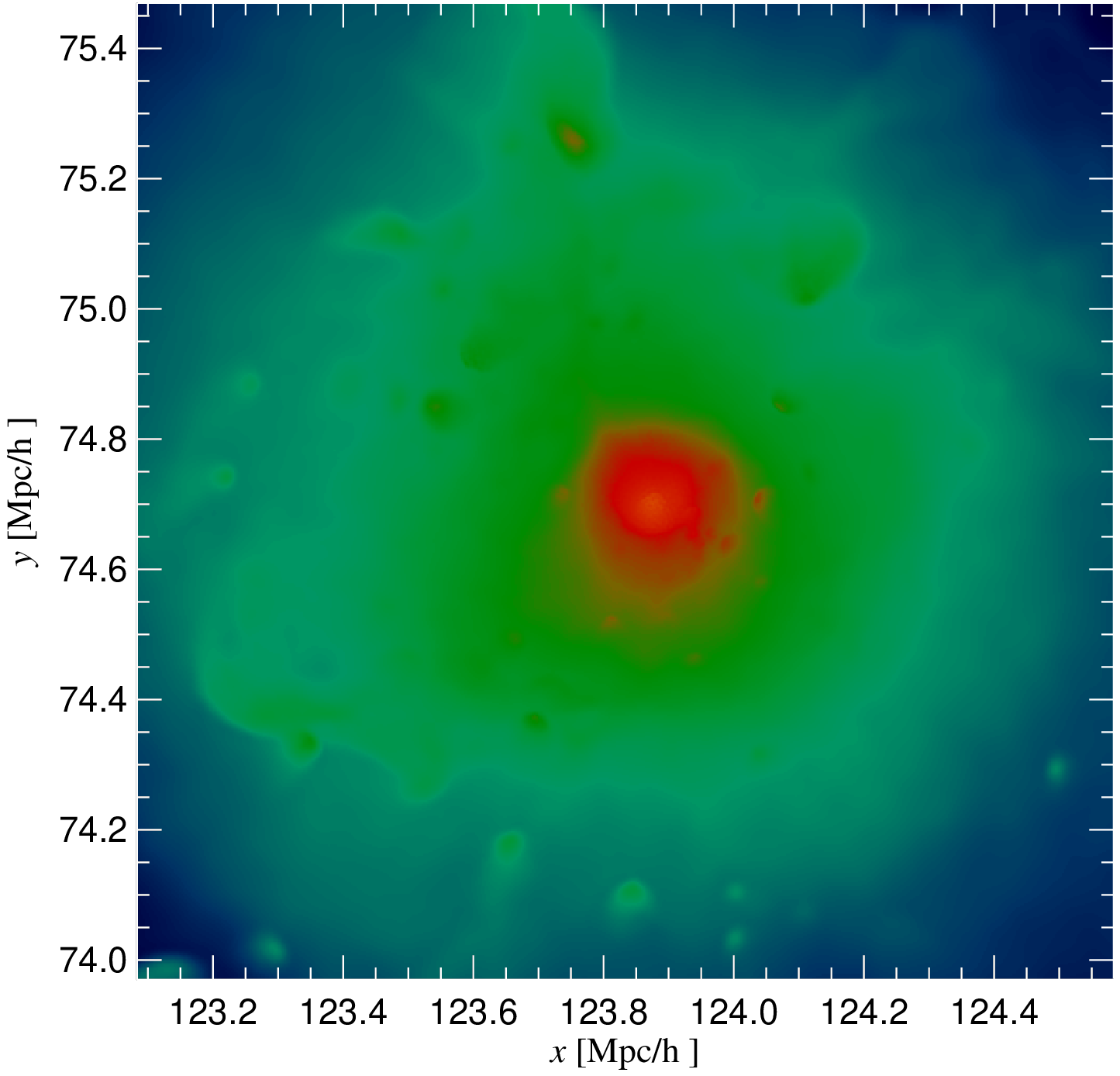}
    \includegraphics[width=8cm, clip, trim = 0cm 1.2cm 0cm 0cm]{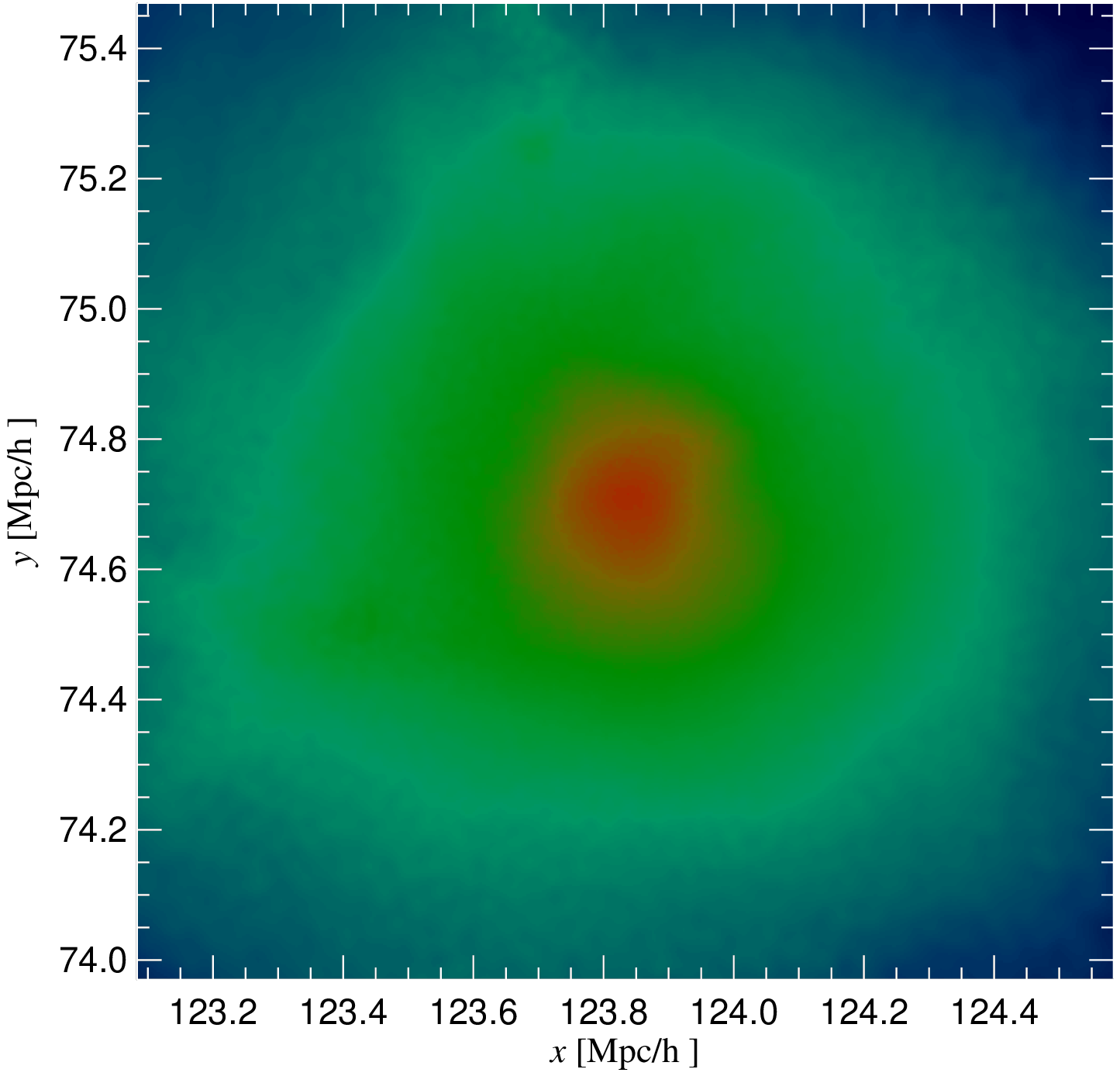}}
  \caption{Projected gas density maps of the $\times$ 128 SPH and SPHS runs (left and right 
    panels), within cubes of side 5 and 2 $h^{-1} {\rm Mpc}$ (top and bottom panels) 
    centred on the cluster at $z$=0.}
  \label{fig:cluster_core}
\end{figure*}

\paragraph*{Visual Impression}In Figure~\ref{fig:SPHS_vs_SPH} we show the
projected gas density distribution within 15 $h^{-1} \rm Mpc$ cubes centred 
on the cluster at $z$=0. The top (bottom) panels correspond to the SPH (SPHS) 
runs, while the left (right) hand panels correspond to the the results of the 
$\times$ 32 and $\times$ 128 (cf. Table~\ref{tab:sim_props}). 

There are several points worthy of note in this figure. First, the 
large-scale spatial distibution of gas is in very good agreement between 
runs -- the cluster resides at the intersection of several filaments, 
and the positions and spatial extents of massive gas clumps are 
consistent across schemes and mass resolutions. Second, and in contrast 
to first point, the spatial distribution of gas on small scales is 
noticeably different between SPH and SPHS -- the number of small dense 
knots in the SPH run is greater than in the SPHS run, and gas clumps are 
more diffuse and extended in the SPHS runs. Third, increasing mass 
resolution has a more striking effect in the SPH runs, with the number 
of dense knots increasing in proportion to the increase in mass 
resolution, whereas this is less obvious in the SPHS runs.

We quantify these second and third points by carrying out a 
friends-of-friends analsysis of the gas density field in the high 
resolution region, adopting a linking length of $b$=0.2 times the mean 
inter-particle separation to compare both the abundance and diffuse 
nature of the gas clumps that form. There are comparable numbers of 
FOF-identified clumps in the SPH and SPHS $\times$8 run, with the number 
of clumps declining as $\sim N_{\rm FOF}^{-0.5}$ for clumps containing 
in excess of $N_{\rm FOF} \sim 50$ particles. For the $\times$ 32 runs, 
we identify $\sim$2-3 times as many gas clumps containing in excess of 
50 particles in the SPH run when compared to the SPHS run. For $N_{\rm 
FOF} \gtrsim$50 particles, there is good agreement between the numbers 
of SPHS clumps in the $\times$8, 32 and 64 runs, with $\sim N_{\rm 
FOF}^{-0.5}$ for increasing $N_{\rm FOF}$, whereas the numbers of SPH 
clumps increases with increasing resolution

\begin{table}
  \begin{center}
    \caption{\textbf{Cluster Properties at $z$=0.}. $M_{\rm vir}$ is the 
      virial mass of the halo, expressed in units of 
      $10^{15} h^{-1} \rm M_{\odot}$, assuming $\Delta_{\rm vir}$=200; 
      $R_{\rm vir}$ is the virial radius, in units of $h^{-1} \rm Mpc$; 
      $N_{\rm tot}$ and $N_{\rm gas}$ are the total number of particles (i.e.
      gas and dark matter) and the number of gas particles within $R_{\rm vir}$;
      and $f_{\rm gas}$ is the baryon fraction 
      $f_{\rm bar} = M_{\rm vir,gas}/M_{\rm vir}$.}
    \vspace*{0.3 cm}
    
    \begin{tabular}{lccccc}
      \hline
      & $M_{\rm vir}$ & $R_{\rm vir}$ & $N_{\rm DM}$&$N_{\rm gas}$ & $f_{\rm bar}$ \\
      & {\scriptsize $10^{15} h^{-1} {\rm M}_{\odot}$}  & {\scriptsize$h^{-1} \rm Mpc$} && & \\
      
      \hline 
      SPH            &        &        &          &       &       \\
      &        &        &         &        &       \\
      $\times$   2   & 6.05 & 1.375 &   26430&12427 & 0.1201 \\
      $\times$   8   & 6.01 & 1.372 &  103897&48079 & 0.1170 \\   
      $\times$  32   & 6.15 & 1.383 &  424138&195903 & 0.1166 \\
      $\times$  64   & 6.23 & 1.389 &  856379&392356 & 0.1151 \\  
      $\times$ 128   & 6.21 & 1.388 & 1716394&791771 & 0.1164 \\  
      $\times$ 256   & 6.21 & 1.387 & 3425723&1581677 & 0.1166 \\
      
      \hline 
      SPHS           &        &        &                  &       &\\
      &        &        &         &                &\\
      $\times$   2   & 6.03 & 1.374 &   26818&12938 & 0.1254 \\   
      $\times$   8   & 5.96 & 1.369 &  107209&52556 & 0.1289 \\   
      $\times$  32   & 6.20 & 1.387 &  445748&218686 & 0.1290 \\  
      $\times$  64   & 6.26 & 1.391 &  898319&438296 & 0.1278 \\  
      $\times$ 128   & 6.20 & 1.386 & 1785832&876886 & 0.1292 \\
      $\times$ 256   & 6.24 & 1.389 & 3578077&1748769 & 0.1282 \\
      \hline
      \hline 
      AMR          &        &        &         &                &\\
      &        &        &         &                &\\
      128   & 4.25 & 1.222 &   25786&- & 0.1397 \\   
      256   & 4.35 & 1.232 &  210674&- & 0.1424 \\   
      \hline      
    \end{tabular}
  \end{center}
  \label{tab:sim_props}
\end{table}

In Figure \ref{fig:cluster_core} we focus on the inner 5 $h^{-1} \rm Mpc$ 
(middle panels) and 2 $h^{-1} \rm Mpc$ (bottom panels) in the SPH and SPHS 
$\times$ 128 runs (left and right panels respectively). Qualitatively we see
evidence that the projected gas density in the core of the SPH run is higher 
compared to its SPHS counterpart. The SPH run also contains a number of dense 
knots of substructure, some of which show evidence of stripping, which are not 
apparent in the SPHS run. 

%%% Compare spherically averaged density, temperature and entropy profiles at
%%% z=0.
\begin{figure}
  \includegraphics[width=8cm, clip, trim = 0cm 1.2cm 0cm 0cm]{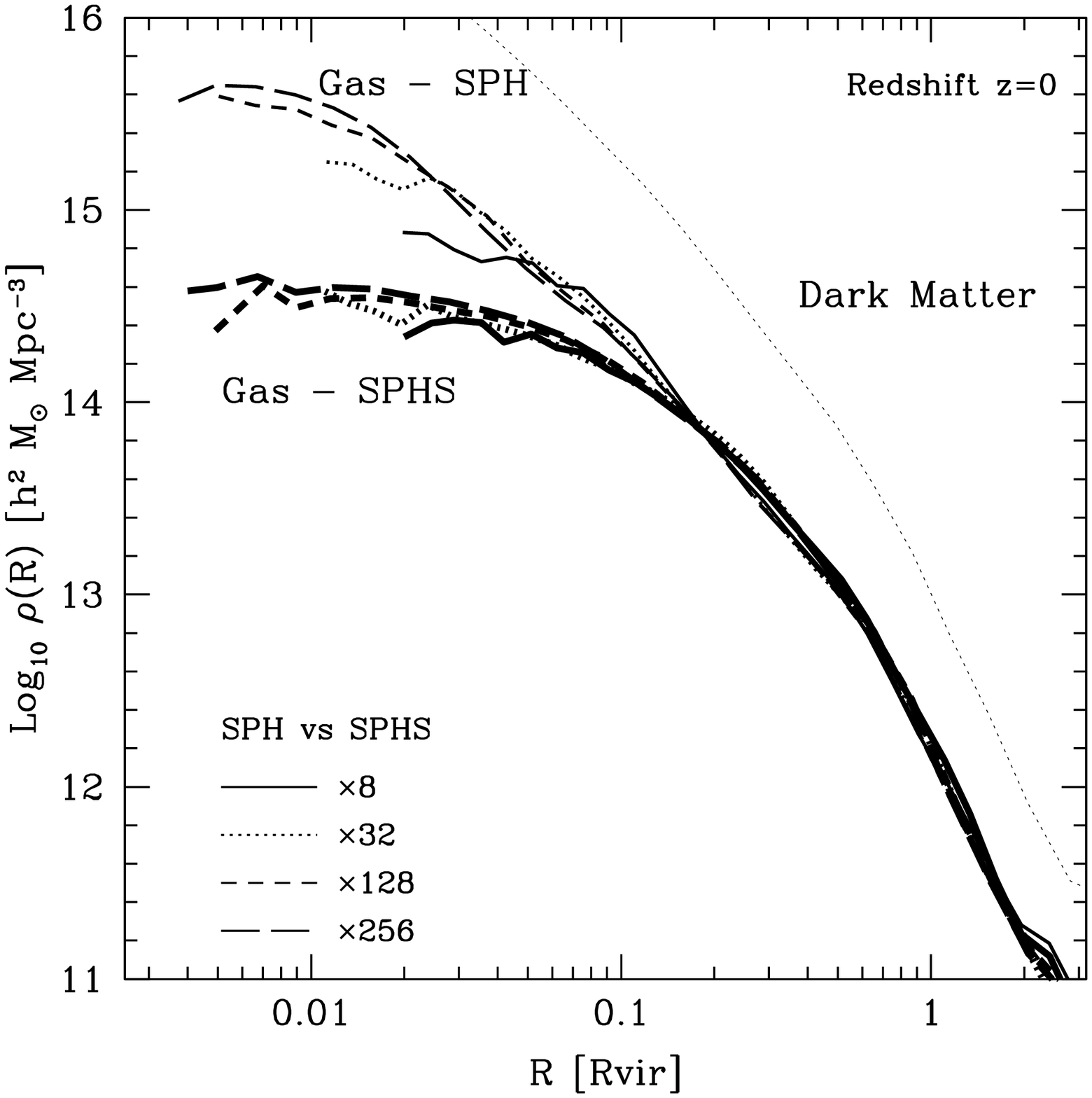}
  \includegraphics[width=8cm, clip, trim = 0cm 1.2cm 0cm 0cm]{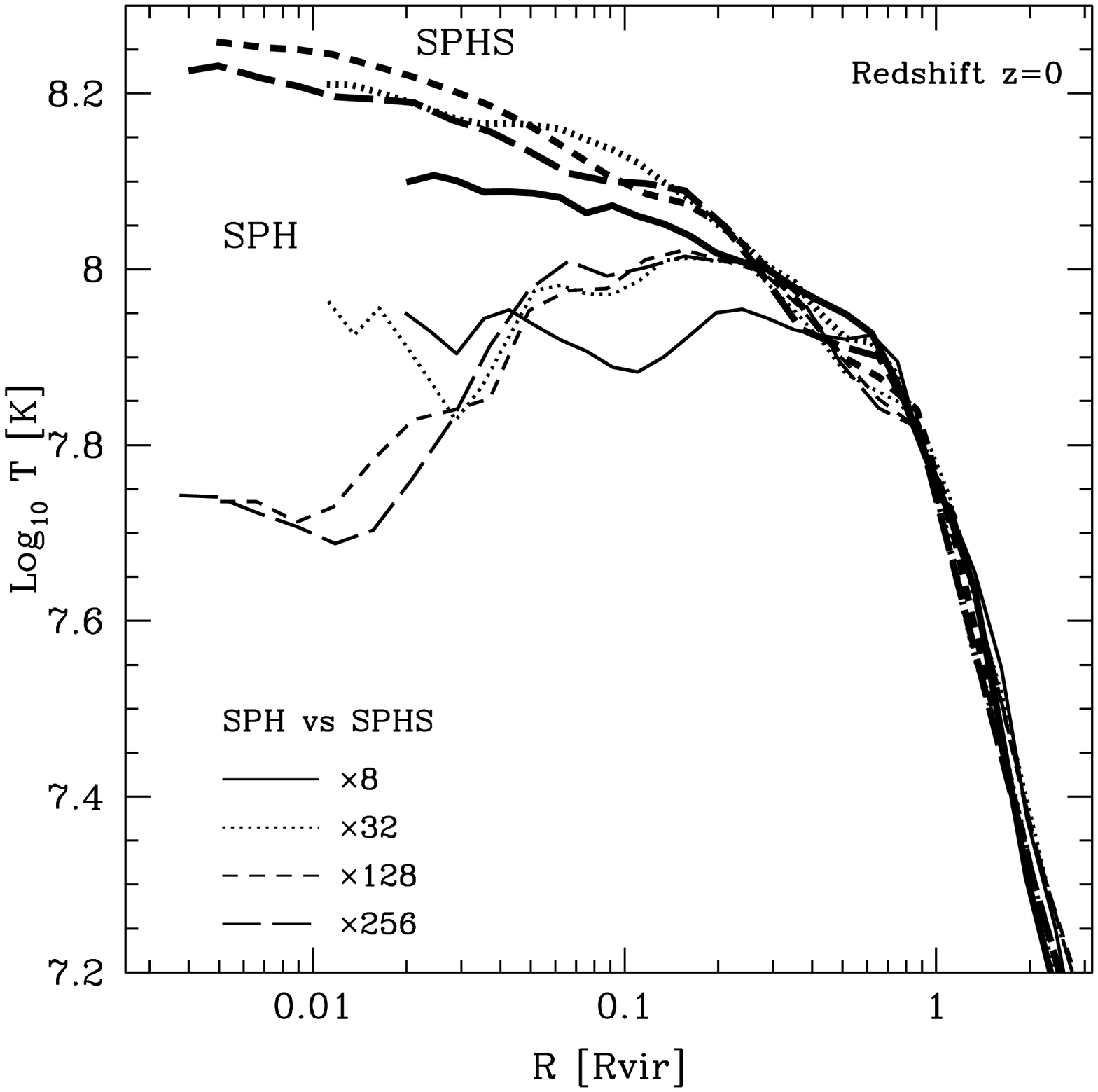}
  \caption{Spherically averaged density (upper panel) and mass-weighted 
    temperature profiles (lower panel) at $z$=0. The heavy (SPHS) and
    light (SPH) solid, dotted, short dashed and long dashed curves correspond
    to the $\times$8, 32, 128 and 256 runs, plotted down to the gravitational 
    softening $\epsilon_{\rm opt}$. The light dotted curve in the upper panel 
    corresponds to the dark matter density profile in the $\times$ 
    256 SPH run.}
  \label{fig:radial_profiles}
\end{figure}

\begin{figure}
  \includegraphics[width=8cm]{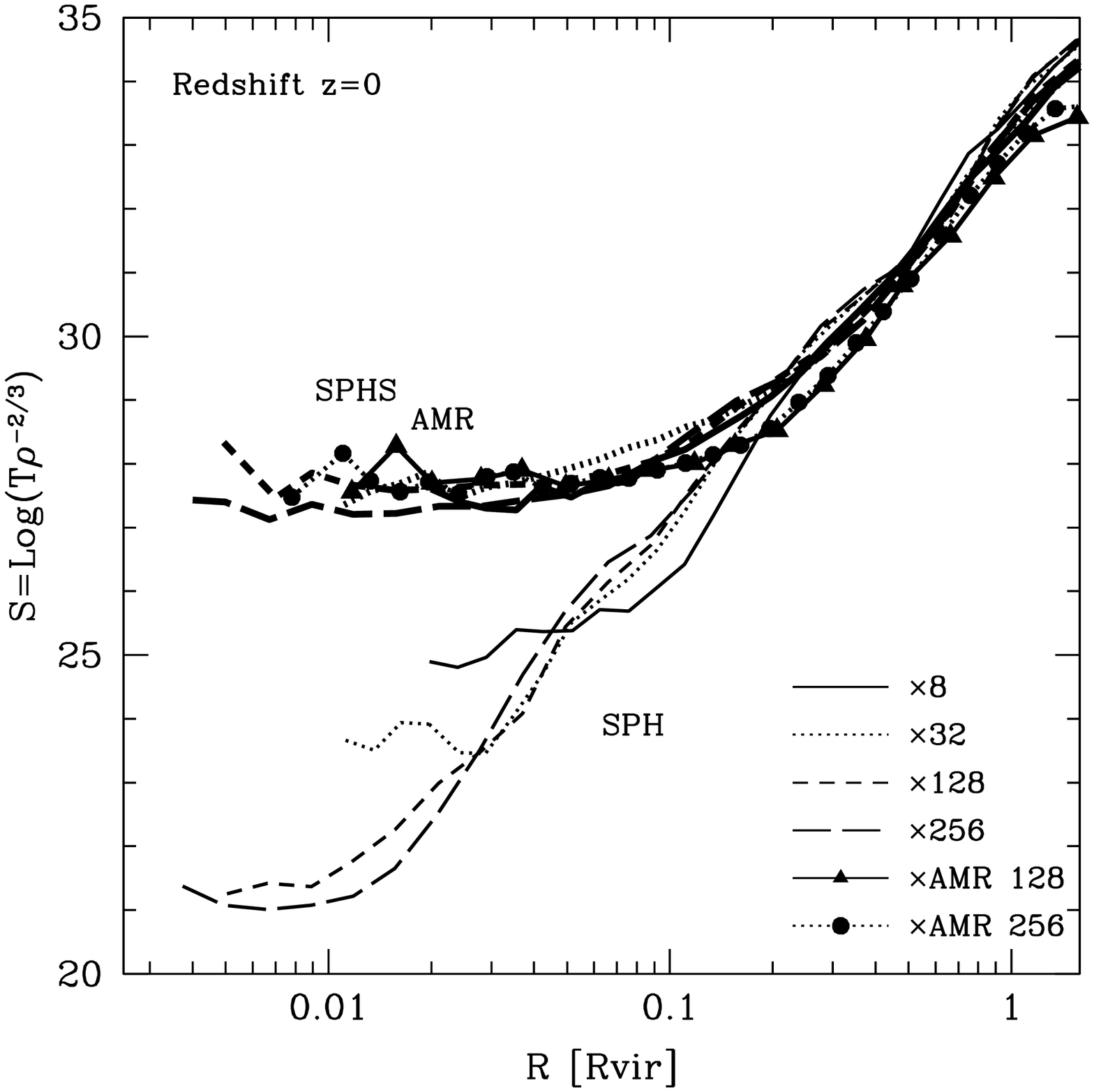}
  \caption{Spherically averaged entropy profiles measured at $z=0$. The 
    heavy (SPHS) and light (SPH) solid, dotted, short dashed and long dashed 
    curves correspond to the $\times$8, 32, 128 and 256 resolution runs, 
    plotted down to the gravitational softening $\epsilon_{\rm opt}$. Overplotted 
    are the results of the AMR 128 and 256 runs -- heavy solid (dotted) curves 
    connecting filled triangles (circles). See text for further discussion.}
  \label{fig:entropy_profiles}
\end{figure}

\paragraph*{Spherically Averaged Profiles} We make these observations 
more precise in Figure~\ref{fig:radial_profiles} and 
\ref{fig:entropy_profiles}. In Figure~\ref{fig:radial_profiles} we plot 
the spherically averaged radial gas density (upper panel) and 
temperature profiles (lower panel) for the $\times$8, 32, 128 and 256
resolution runs, including only bins which contain in excess of 10 gas 
particles (this affects only the innermost bins) down to the gravitational
softening scale $\epsilon_{\rm opt}$. For comparison we plot 
also the dark matter density profile down to the converged radius for 
the $\times$ 256 SPH run. The spherically averaged gas density is 
systematically lower in the SPHS run when compared to the SPH run at 
fixed cluster-centric radius, with the disparity increasing with 
decreasing radius -- from $\sim$ 0.4 dex in the outer parts $\gtrsim$ 
10\% $R_{\rm vir}$ to $\sim$ 1.4 dex within the central 10\% $R_{\rm 
vir}$, in keeping with our observations in 
Figure~\ref{fig:cluster_core}. The temperature profiles are in 
reasonable agreement at large radii ($\gtrsim$ 20\% $R_{\rm vir}$) for 
well resolved runs ($\times$ 32, 128) but they diverge at small radii by 
as much as 0.4 dex. 

There is good consistency between the gas density profiles in the SPHS 
runs -- to better than 10\% to within 0.1 (0.03) $R_{\rm vir}$ for the 
$\times$8 (32) runs, compared to $\times$128. In contrast, the gas 
density profiles in the SPH runs start to deviate by greater than 10\% 
at 0.2 (0.03) $R_{\rm vir}$ for the $\times$8 (32) runs, compared to 
their $\times$ 128 counterpart. A similar degree of consistency is 
evident in the temperature profiles -- to better than 10\% to within 0.1 
(0.02) $R_{\rm vir}$ for the $\times$8 (32) SPHS runs with respect to 
the fiducial $\times$ 128 SPHS run, and to better than 10\% to within 
0.3 (0.3) $R_{\rm vir}$ for the $\times$8 (32) SPH runs with respect to 
the fiducial $\times$ 128 SPH run. 

These results show that SPH produces cluster cores that are have higher 
central densities and lower central temperatures than their counterparts 
in SPHS, and so we expect systematically lower entropies in SPH, 
according to according to Equation~(\ref{eq:entropy}). This is evident 
in Figure~\ref{fig:entropy_profiles}, in which we compare entropy 
profiles at $z$=0 in the SPH and SPHS runs -- as well as one of the AMR runs,
which we shall discuss below. The entropy profiles in the 
SPH runs are consistent with those report by previous studies, declining 
with decreasing cluster-centric radius. There is no obvious convergence 
with increasing mass resolution -- the entropy continues to reach 
smaller values as the mass and spatial resolution improves. In contrast, 
the entropy profiles plateau to a well-defined value in the SPHS runs 
and there is excellent agreement between the different resolution runs
for $R \gtrsim 0.02 R_{\rm vir}$. Physically this is not unreasonable -- 
if the cluster is in approximate hydrostatic equilirbrium, as we might
expect during a period of quasi-dynamical equilibrium, we would 
expect the central gas density to plateau for both isothermal and
polytropic equations of state \citep[cf.][]{1998ApJ...497..555M,
2001MNRAS.327.1353K}, with the consequence that the central entropy 
profile should also plateau.

\begin{figure*}
  \centerline{\includegraphics[width=6cm]{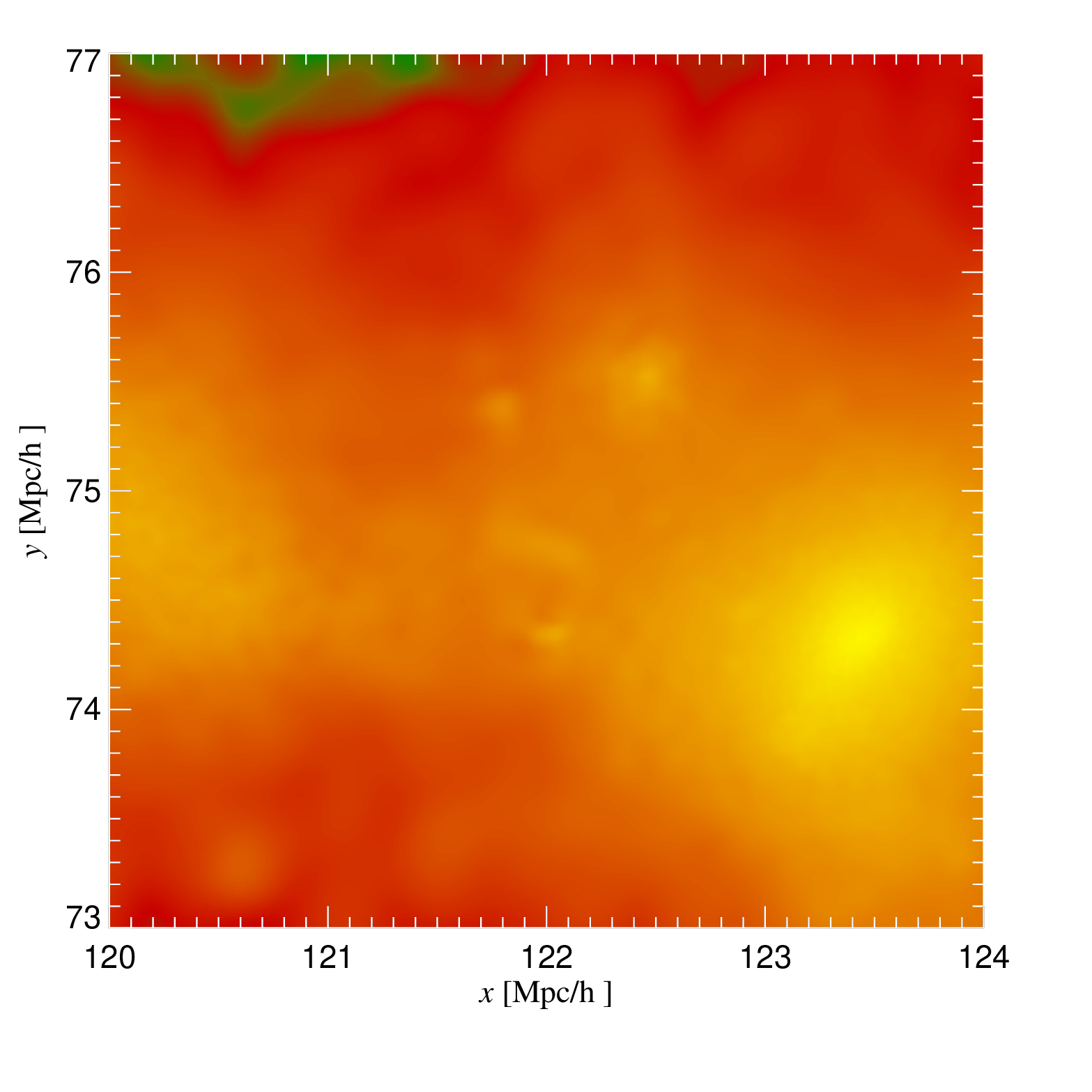}
    \includegraphics[width=6cm]{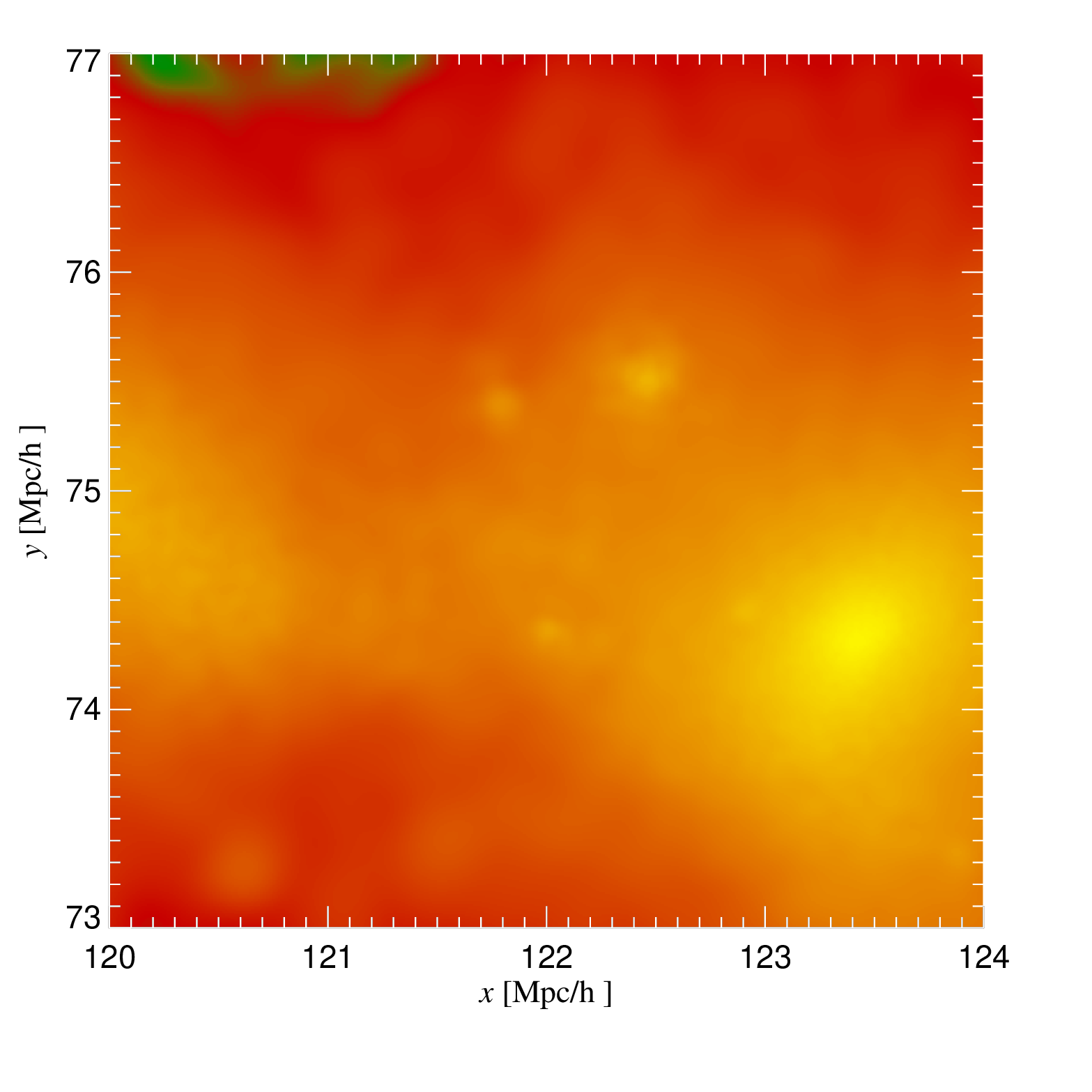}
    \includegraphics[width=6cm]{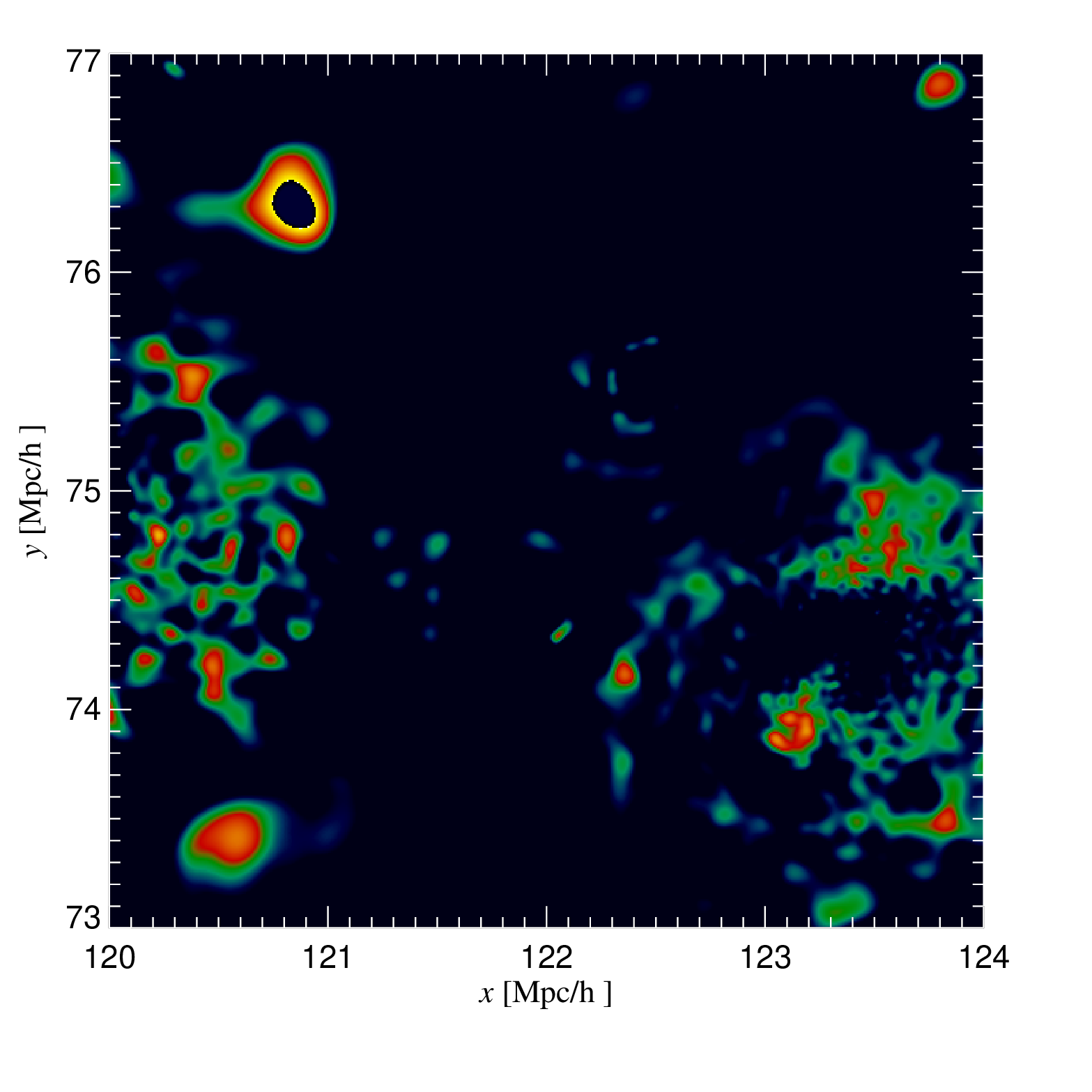}}
  \centerline{\includegraphics[width=6cm]{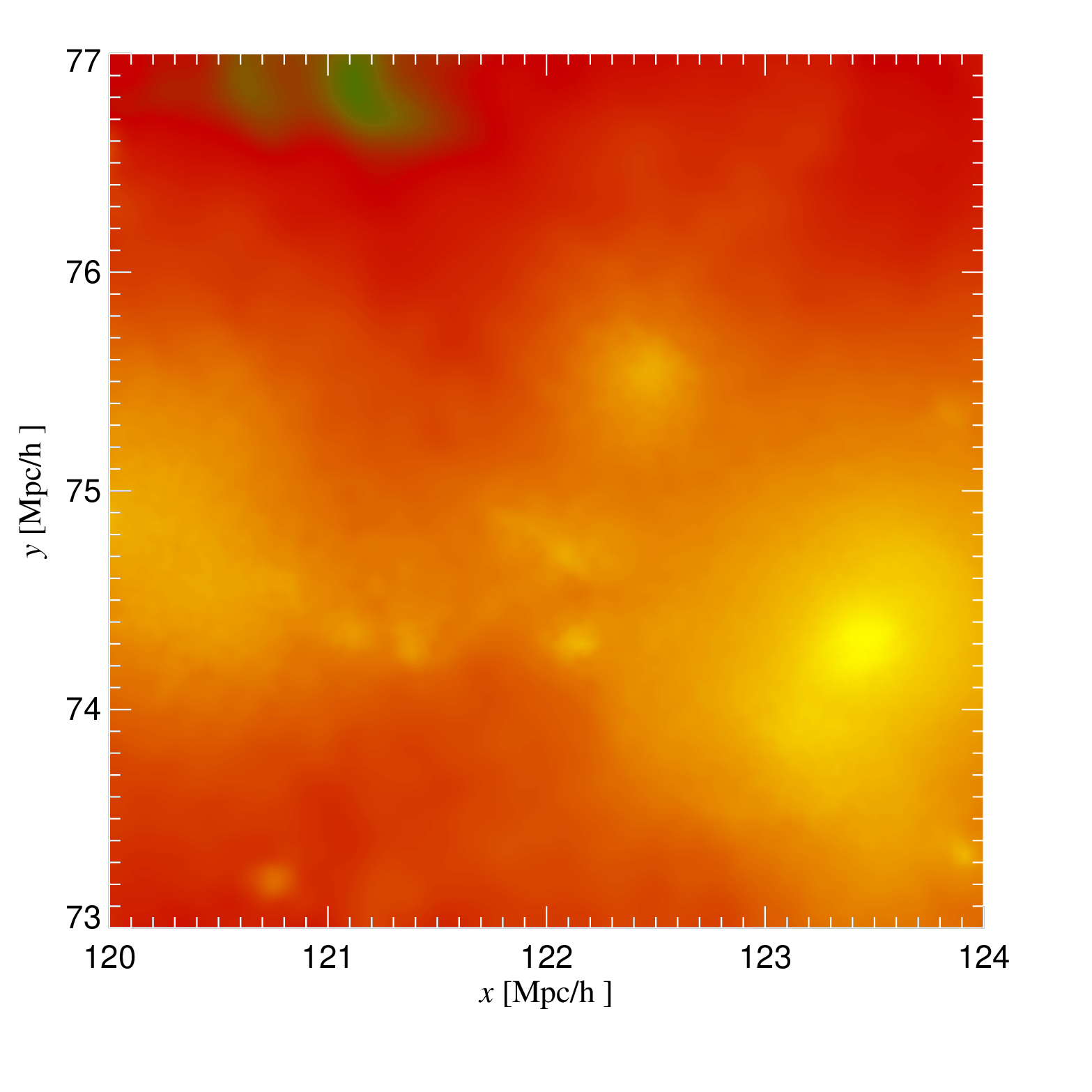}
    \includegraphics[width=6cm]{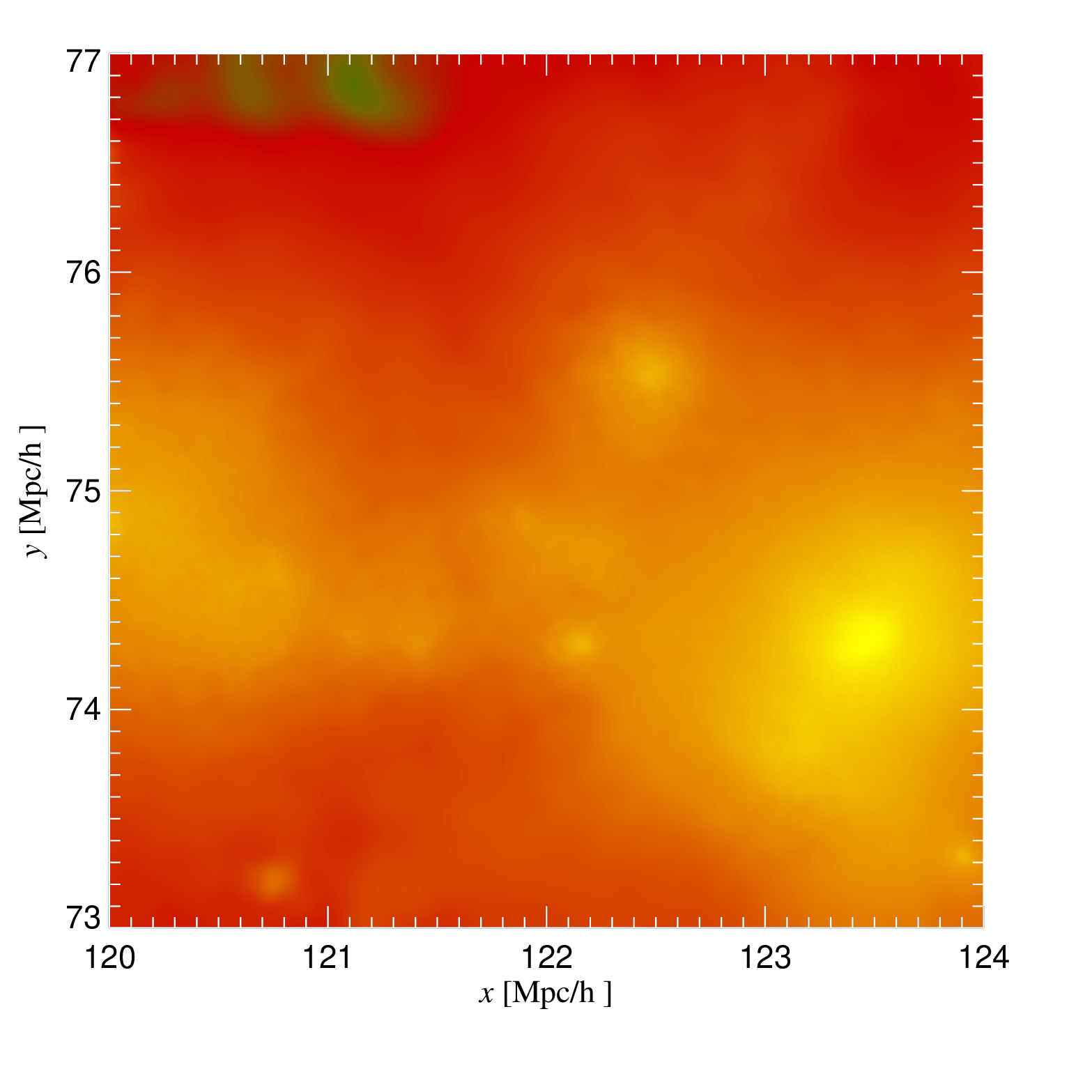}
    \includegraphics[width=6cm]{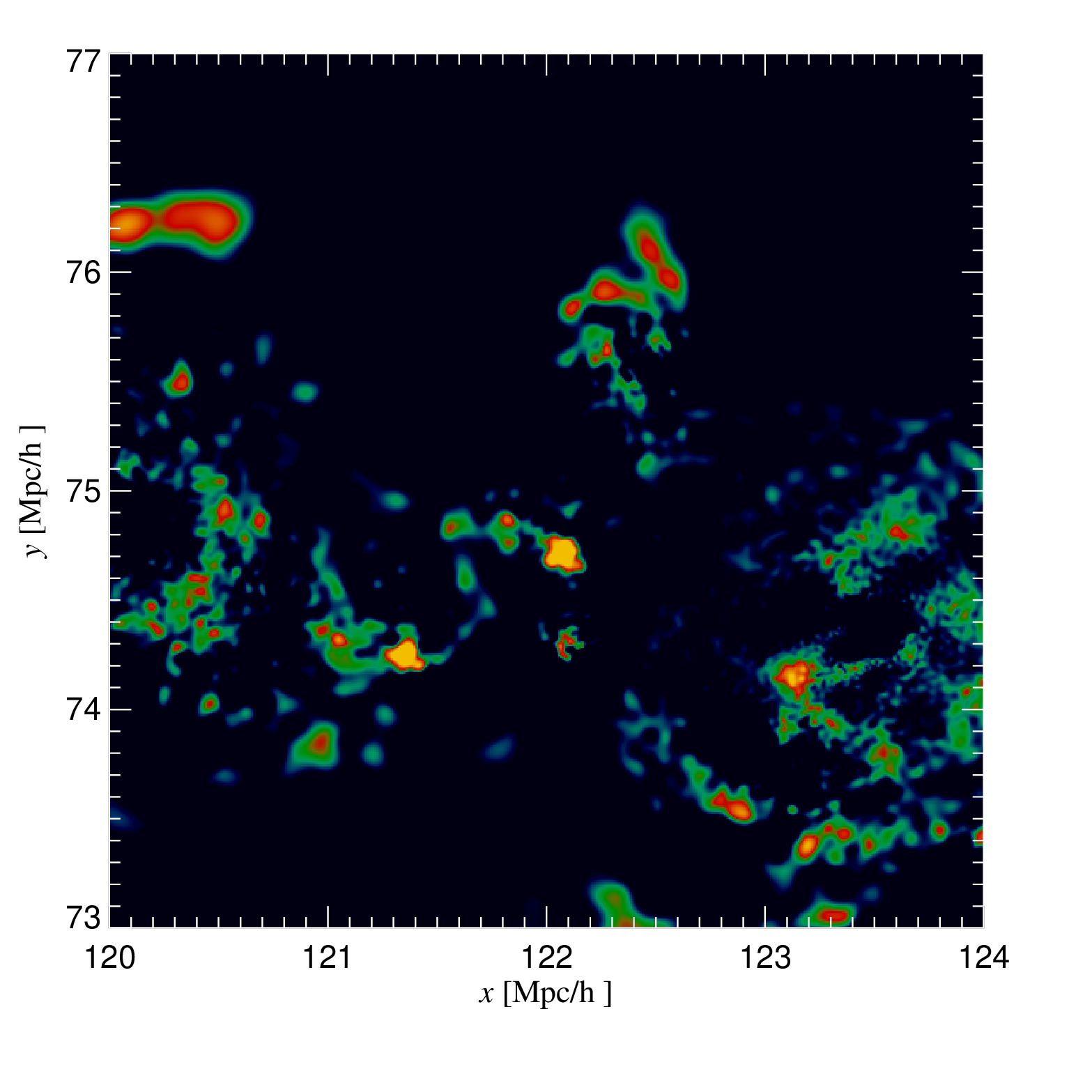}}
  \caption{\textbf{Visual Impression.} Maps of the projected gas density 
    in a cube of side 4 $h^{-1} {\rm Mpc}$ centred on the outskirts of the 
    cluster at $z$=1 in the $\times$8 and $\times$32 SPHS runs (top and bottom 
    panels). Left and middle panels show results for runs in which 
    $\alpha_{\rm max}$=1 and $\alpha_{\rm max}$=5 respectively; right panels 
    plot $\kappa(x,y)=[P_1-P_5]/[P_1+P_5]$, where 
    $P_{\alpha_{\rm max}}$ is the value of the pixel at $(x,y)$ in the run with
    given $\alpha_{\rm max}$. }
\label{fig:SPHS_alphamax}
\end{figure*}

\paragraph*{Sensitivity to Artificial Bulk Dissipation Constant 
$\alpha_{\rm max}$.} SPHS invokes numerical dissipation in converging 
fluid flows to suppress multi-valued fluid quantities (e.g. pressure), 
which lead to large numerical errors. This is controlled by $\alpha_{\rm 
max}$, whose default value is 1 and which in the limit of $\alpha_{\rm 
max} \rightarrow 0$ should produce results that are more similar to 
classic SPH (though not identical, since SPHS has improved force accuracy 
as compared to classic SPH; see \S\ref{sec:sims}). 
Figure~\ref{fig:SPHS_alphamax} provides a visual impression of the SPHS 
density field, centred on the outskirts of the cluster at $z$=1, in 
$\times$ 8 and 32 runs (top and bottom panels) assuming values of 
$\alpha_{\rm max}$=1 and $\alpha_{\rm max}$=5 (left and right panels 
respectively). We expect $\alpha_{\rm max}$=5 runs to be more dissipative
than $\alpha_{\rm max}$=1, and we expect dissipation to shift to smaller 
scales with increasing mass resolution. These effects, although subtle, 
are borne out in the left and middle panels of Figure~\ref{fig:SPHS_alphamax} 
(focus, for example, on the small substructure in the $\times$32 simulations 
at $\sim[122.2,74.8]$\,Mpc/h).
We quantify this further in the right panels, by plotting the logarithm of:
\[
\kappa(x,y) = \frac{P_1-P_5}{P_1+P_5}
\]
where $P_{\alpha_{\rm max}}$ is the value of the pixel at coordinate $(x,y)$ in the
run of given $\alpha_{\rm max}$. Notice that the differences between the 
$\alpha_{\rm max}$=1 and $\alpha_{\rm max}$=5 simulations shift to smaller scales
 with increasing resolution.

\begin{figure}
  \includegraphics[width=8cm]{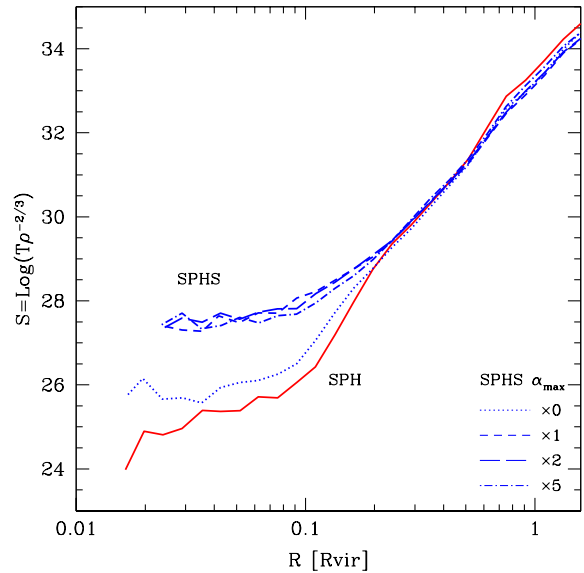}
  \caption{\textbf{Sensitivity to Artificial Bulk Dissipation Constant $\alpha_{\rm max}$.} 
    Spherically averaged entropy profiles assuming dissipation is switched off 
    (i.e. $\alpha_{\rm max}$=0; dotted-dashed curve); switched on and set to its default 
    value (i.e. $\alpha_{\rm max}$=1; long dashed); and switched on and set to 
    $\alpha_{\rm max}$=2 and 5 (short dashed and dotted respectively). For comparison
    we show also the profile from the corresponding classic SPH run (solid curve).}
  \label{fig:Entropy_Profiles_Diss}
\end{figure}

Figure~\ref{fig:Entropy_Profiles_Diss} 
shows the spherically averaged entropy profiles at $z=0$ in four runs at 
$\times$8 resolution -- one in which numerical dissipation is switched off 
(i.e. $\alpha_{\rm max}=0$), one set to its default value (i.e. $\alpha_{\rm 
max}=1$), and two with $\alpha_{\rm max}=2$ and $5$. For $\alpha_{\rm max} 
\ge 1$ the profiles are converged; for $\alpha_{\rm max}=0$, the profile 
is declining with decreasing radius, albeit less sharply than the 
entropy profile from the corresponding $\times$8 SPH run. These results 
demonstrate that the differences between SPH and SPHS are largely driven 
by the numerical dissipation implemented in SPHS, rather than the improved 
force accuracy. 

\subsection{Comparison with AMR}

\begin{figure}
  \centerline{
    \includegraphics[width=8cm, clip, trim = 0cm 1.2cm 0cm 0cm]{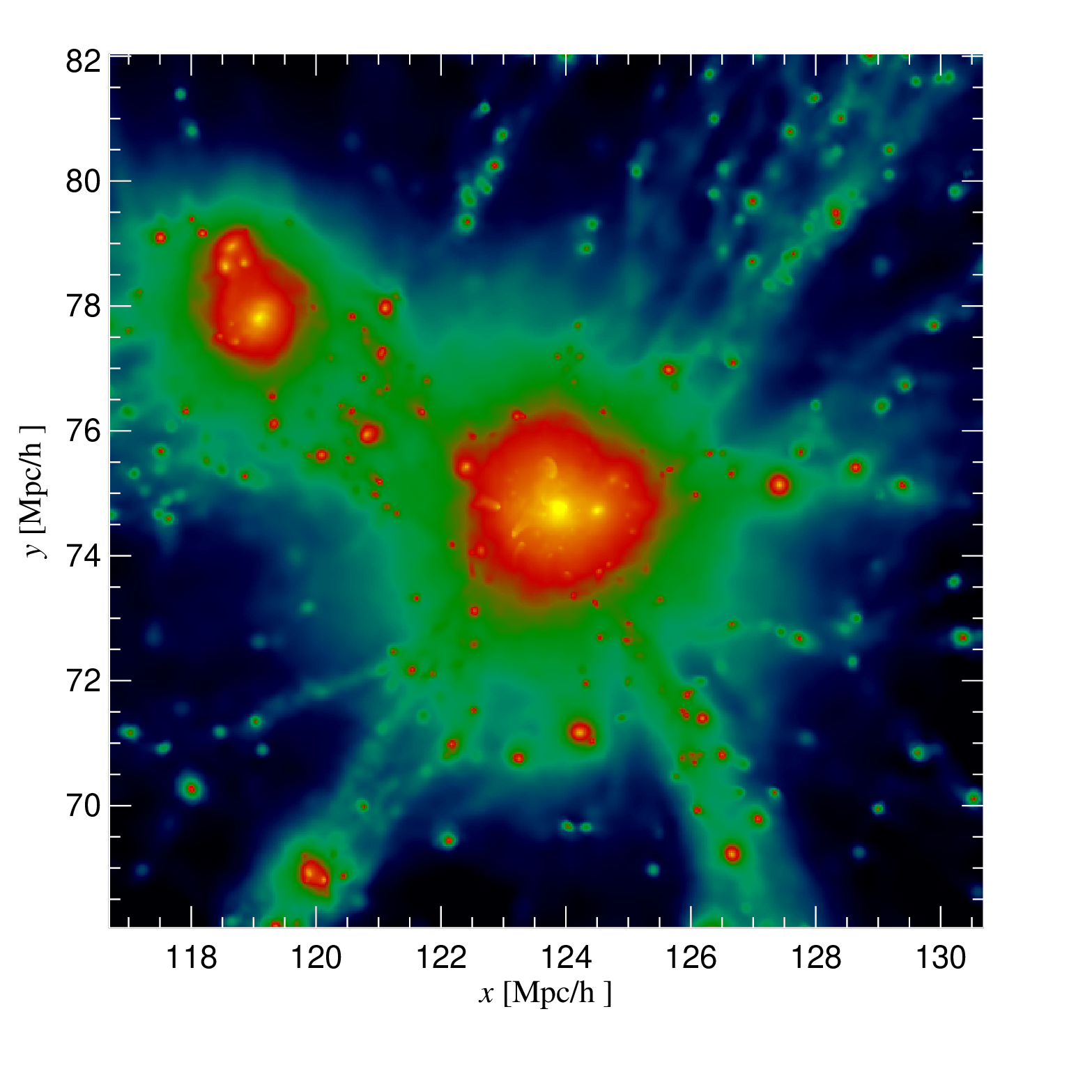}}
  \centerline{
    \includegraphics[width=8cm, clip, trim = 0cm 1.2cm 0cm 0cm]{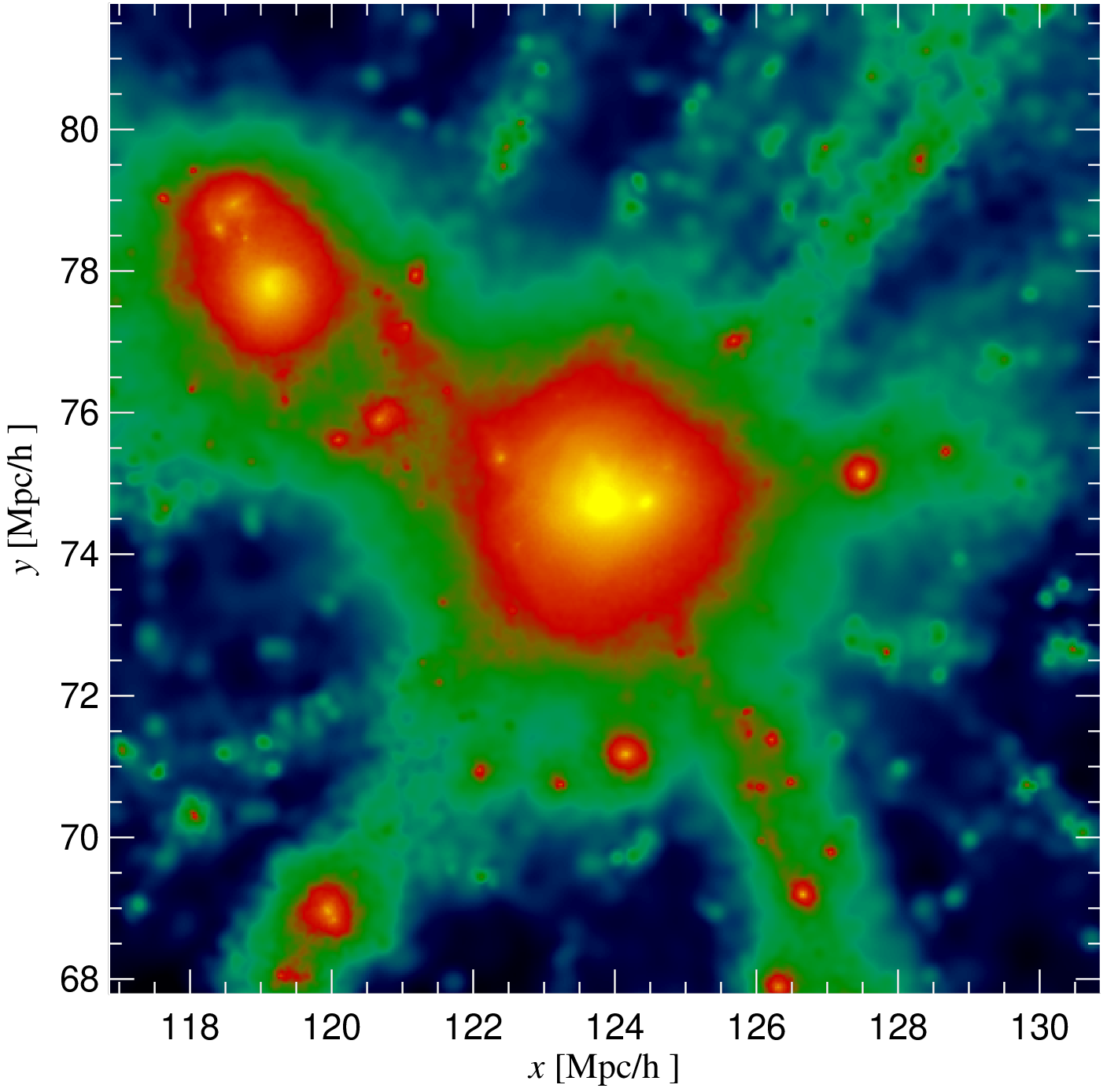}}
  \centerline{
    \includegraphics[width=8cm, clip, trim = 0cm 1.2cm 0cm 0cm]{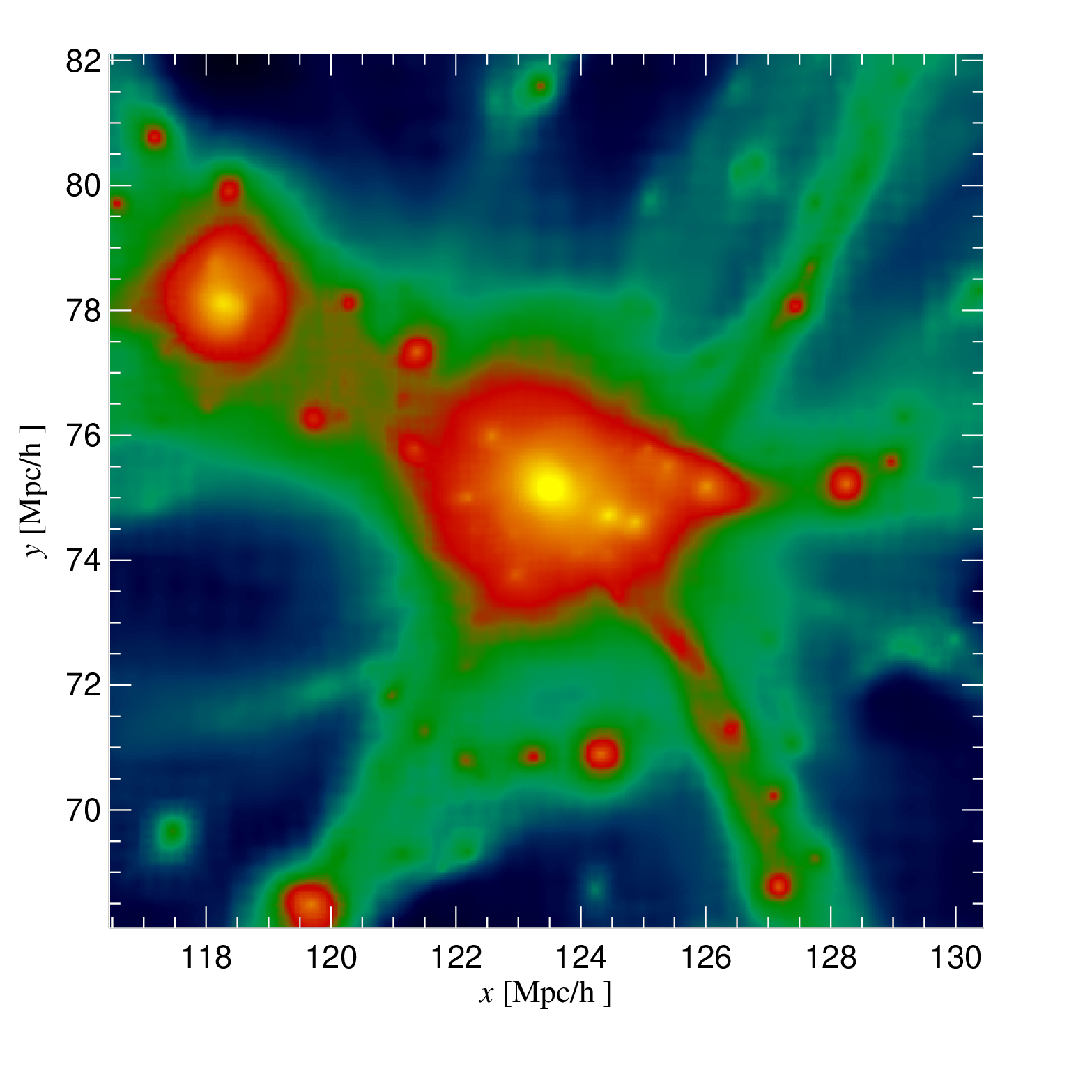}}
  \caption{\textbf{Code Comparison: Visual Impression} Projected gas density maps in the 
      SPH, SPHS and AMR runs (from top to bottom) within a 15 $h^{-1} {\rm Mpc}$ cube 
      centred on the cluster at $z=0$.}
  \label{fig:ClusterCoreComparison}
\end{figure}

We show projected density maps of the gas within a cube of side 15 
$h^{-1} {\rm Mpc}$ centred on the cluster at $z$=0 in the SPH, SPHS and 
AMR runs in Figure~\ref{fig:ClusterCoreComparison}. The large-scale 
spatial distribution of gas is similar across the runs -- the cluster 
forms at the intersection of several filaments that are funneling lower 
mass systems towards it. It is evident from these density maps, and from 
comparison of cluster virial masses and merging histories, that we have 
captured the cluster at a slightly earlier stage of its evolution in the 
AMR run compared to the SPH and SPHS runs -- the cluster has yet to 
merge with the complex of lower mass structures at projected position 
$(x,y) \sim (125,75) h^{-1} \rm Mpc$ in the AMR run, whereas this merger 
that occurred at $z \sim$ 0.14 in the SPH and SPHS runs. This difference 
in timing reflects structural differences in the low resolution mass 
distribution between the \texttt{GADGET} and \texttt{RAMSES} runs, which 
in turn affects the large-scale gravitational field and consequently 
halo dynamics. 

Despite these differences, we find very good consistency between entropy 
profiles in the SPHS and AMR runs, as shown in Figure~\ref{fig:entropy_profiles}.
Recall that the SPHS and SPH results -- for the $\times$8, 32, 128 and 256 resolutions 
-- are indicated by heavy and light curves respectively, while the results of the 
AMR 128 and 256 runs are heavy solid (dotted) curves connecting filled 
triangles (circles). There is excellent agreement between the two AMR runs,
while the level of agreement between the SPHS and AMR runs is impressive -- it 
is as good as the scatter in the central entropy profile across the different 
resolution SPHS runs. 

This scatter in central entropy in the SPHS and AMR runs is to be expected -- we 
are modelling a chaotic non-linear system and so as our mass and force resolution 
increases so too does our power to resolve smaller scale perturbations, which will 
be imprinted on the central entropy profile at later times. However, we do not 
expect significant changes in the central entropy in either the SPHS or AMR runs 
as we go to higher mass and force resolution; we may resolve smaller-scale perturbations 
and form lower mass substructures, but these substructures will find it as difficult, 
if not more so, to retain their gas, and their lower masses imply that will have 
correspondingly longer dynamical friction and (consequently) merging timescales.

We can see why by noting that the ram pressure acting on these substructures as they 
pass close to the cluster core is of order $\rho_{\rm cl}\sigma_{\rm cl}^2$, where
$\rho_{\rm cl}$ is the central density of the cluster and $\sigma_{\rm cl}$ is the 
cluster's velocity dispersion, and this ram pressure will be effective provided 
$\rho_{\rm cl}\sigma_{\rm cl}^2 \gtrsim \rho_{\rm gas,sub}\sigma_{\rm sub}^2$, where
$\rho_{\rm gas,sub}$ is the gas density within the substructure and $\sigma_{\rm sub}^2$
is the substructure's velocity dispersion. Lower mass substructures will be more 
concentrated and so will have higher gas densities (i.e. $\rho_{\rm gas,sub}>\rho_{\rm cl}$),
but the rate of increase of concentration with decreasing mass is a very weak
function of mass \citep[e.g. $\sim M^{-0.1}$, cf.][]{2007MNRAS.381.1450N} and so, 
in the absence of cooling, the decrease in $\sigma$ dominates; lower mass substructures become
progressively more susceptible to ram pressure stripping.

If anything, the presence of this population of substructures will 
be more of a blight for the classic SPH runs, because they are more likely to retain 
their gas and the passage of these cooler clumps through the cluster core will lead 
to more frequent shocking and stirring, leading to fluctuations in the central entropy 
that may not be evident in lower resolution runs.

%%% Examine redshift evolution of the entropy profile.

\subsection{Redshift Evolution} 

\begin{figure}
    \includegraphics[width=8cm]{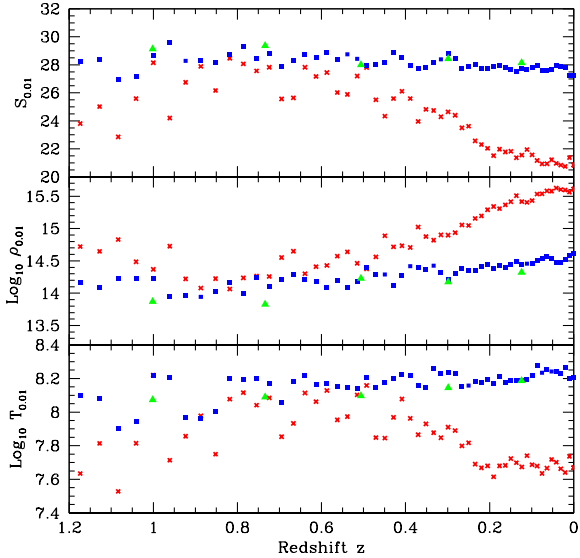}
  \caption{Redshift variation of the spherically averaged entropy $S_{0.01}$,
    density $\rho_{0.01}$ and temperature $T_{0.01}$ (upper, middle and lower panels)
    measured at a fiducial cluster-centric radius of 0.01 $R_{\rm vir}$. Crosses, filled 
    squares and filled triangles correspond to SPH, SPHS and AMR results.}
  \label{fig:Redshift_Entropy_Profiles_SPH_vs_SPHS}
\end{figure}

So far, we have compared cluster properties 
at $z=0$ in the SPH, SPHS and AMR runs. We now consider cluster properties
at earlier times, whose evolution we distill in 
Figure~\ref{fig:Redshift_Entropy_Profiles_SPH_vs_SPHS}. Here we show how 
the spherically averaged estimates of the entropy, density and 
temperature (top, middle and bottom panels), measured at a fiducial radius
$R_{0.01}$=0.01$R_{\rm vir}$, have varied with redshift since $z\sim 1$. Results 
from the $\times$ 256 SPH and SPHS runs are indicated by crosses and filled 
squares; filled triangles correspond to the results from the AMR256 run.

\begin{figure}
  \includegraphics[width=8cm]{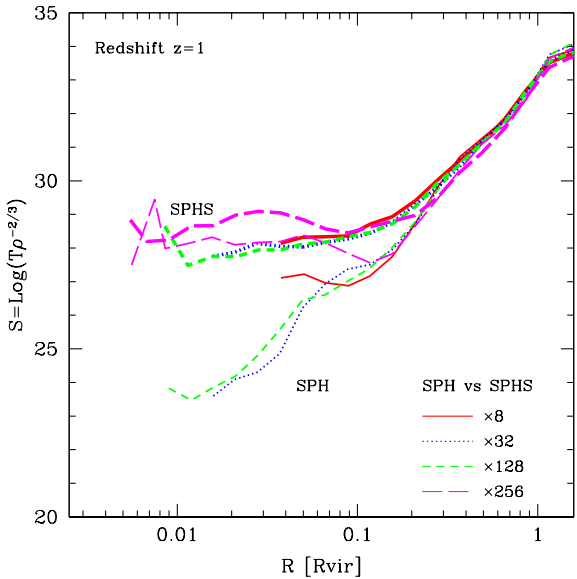}
  \caption{Spherically averaged entropy profiles at $z$=1. As in 
    Figure~\ref{fig:entropy_profiles}, the heavy (SPHS) and light (SPH)
    solid, dotted, short dashed and long dashed curves correspond to 
    the $\times$8, 32, 128 and 256 resolution runs, plotted down to the 
    gravitational softening $\epsilon_{\rm opt}$.}
  \label{fig:Redshift_Profiles}
\end{figure}

This is a revealing figure for a number of reasons. First, it indicates 
that the SPH and SPHS runs produced broadly consistent results at $z \gtrsim 0.6$,
but have diverged since then such that there is a factor of 10 (3) difference 
in the estimated density (temperature) at $z$=0. If we look at the spherically
averaged profiles in detail at, say, $z \sim 1$ (cf. Figure~\ref{fig:Redshift_Profiles}) 
we find that a plateau can form in the SPH entropy profile (cf. the $\times$ 256 run)
and there can be reasonable consistency between the SPH and SPHS profiles, 
but this plateau is a transient feature in the SPH case whereas it is long-lived
in the SPHS case, and it is also resolution dependent (compare the $\times$ 256 
profile to the $\times$ 8, 32 and 128 profiles, which are declining with radius).

This relates to our second observation, which is the relative
stability of the SPH entropy profile compared to the SPH entropy profile; 
since $z=1.2$, $S_{0.01} \sim 28$ in the SPHS run whereas $S_{0.01}$ has fluctuated 
and spanned the range $22 \lesssim S_{0.01} \lesssim 28$. These fluctuations track
the violent assembly history of the cluster; it has a formation 
redshift\footnote{Following convention, we define the formation redshift 
cluster as the redshift at which half its $z=0$ virial mass is in place 
\citep[cf.][and references therein]{power.etal.2012}} of
$z_{\rm form}=0.5$, which is typical for the most massive galaxy clusters, and
it has assembled 70\% of its $z=0$ mass since $z=1$. As noted earlier, a key
difference between the SPH and SPHS runs is the abundance of low mass, 
high density clouds evident in the SPH density field that are not present
in the SPHS density field. These clouds are associated low entropy material in the 
cores of underlying dark matter substructures; as they plunge towards the cluster
centre and merge, gas is stripped and flung outwards, shocking to high temperatures. 
The low entropy material settles in the cluster core and gives rise to the lower 
entropy profile evident at $z=0$, but the shocked gas in the cluster core takes time
to expand and redistribute, stirring the cluster gas in the process. The same dark matter
substructures are evident in the SPHS and AMR runs, but if they are occupied by gas 
it is at a higher entropy and so is more easily stripped by the intra-cluster medium.

Third, we note excellent consistency between the SPHS and AMR runs holds at earlier
times. This is also evident in the entropy profiles (cf. Figure~\ref{fig:Redshift_Profiles}).

\section{Discussion}
\label{sec:discussion}

\subsection{What is the origin of the discrepancy between the classic 
SPH and the AMR results?}\label{sec:sphvamr}

Since the work of \citet{1999ApJ...525..554F}, it has been known that SPH and AMR codes 
produce very different results for the entropy profile of the intracluster medium in 
non-radiative simulations of a massive galaxy cluster. Numerous studies in the literature 
have suggested that the SPH results are flawed 
\citep{2008MNRAS.387..427W,2009MNRAS.395..180M,2011arXiv1109.3468S}, most likely owing to a spurious numerical 
surface tension \citep{2007MNRAS.380..963A}. Indeed, a recent simulation using SPH with 
dissipation in entropy has reported an entropy core more similar to that found in AMR or 
mesh-based simulations \citep{2008MNRAS.387..427W}. However, the amplitude of this core 
was found to be sensitive to both the choice of numerical dissipation parameters and the 
numerical resolution (see \S\ref{sec:dissres} for further discussion of this). In this 
paper, we have tested a new SPH algorithm -- SPHS -- that is designed to converge with 
increasing resolution independently of the choice of dissipation parameters 
\citep{2012MNRAS.422.3037R}; we also present explicit comparisons with an AMR code {\tt 
RAMSES}, similarly to the original study of \citep{1999ApJ...525..554F}. Our suite of 
simulations that explore resolution, dissipation parameter and choice of hydrodynamic 
solver allow us to pin-point the precise reasons for the differences between the SPH and 
AMR simulations.

\begin{figure*}
  \centerline{
    \includegraphics[width=8cm, clip, trim = 0cm 1.2cm 0cm 0cm]{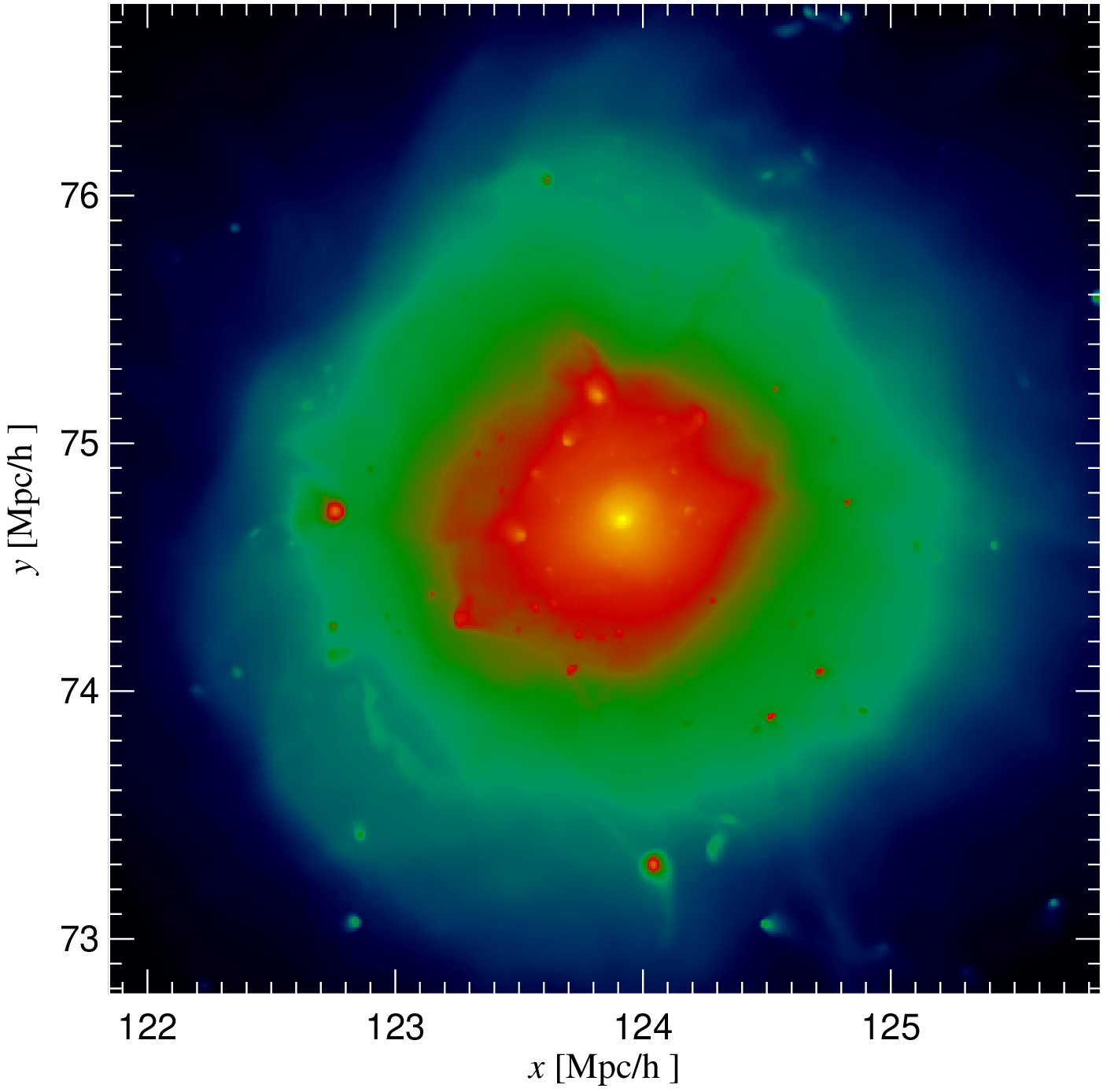}
    \includegraphics[width=8cm, clip, trim = 0cm 1.2cm 0cm 0cm]{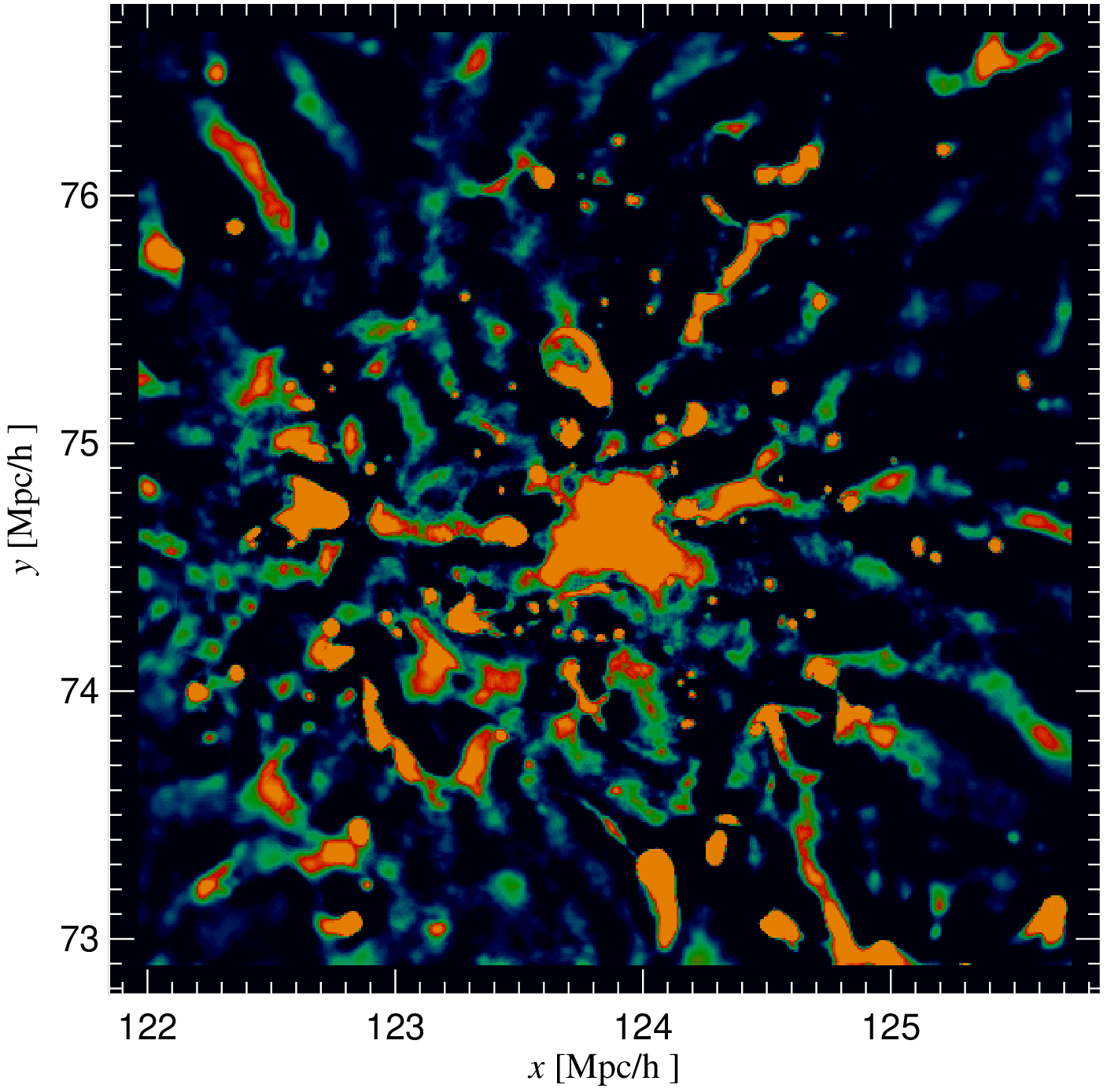}}
  \centerline{
    \includegraphics[width=8cm, clip, trim = 0cm 1.2cm 0cm 0cm]{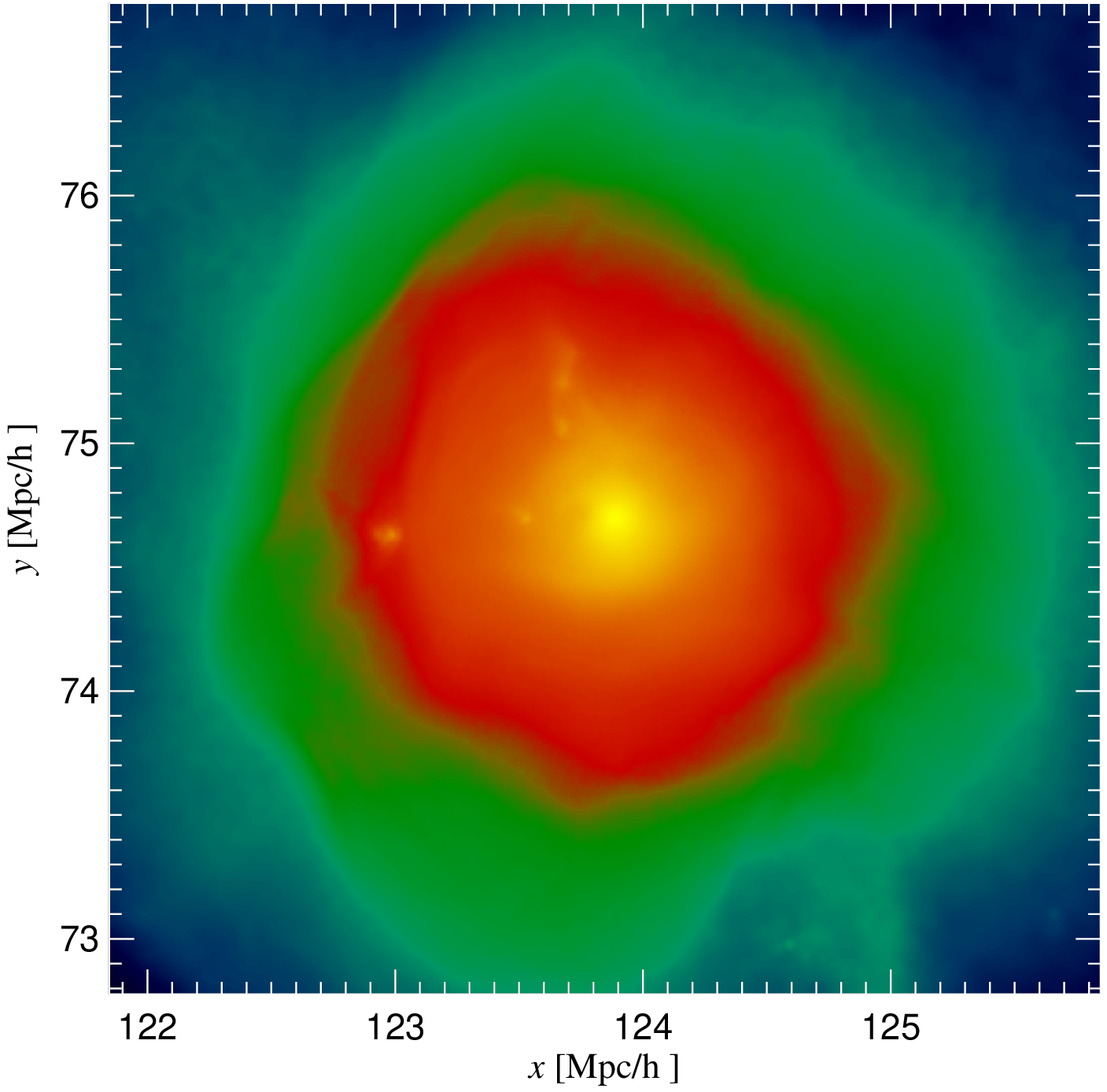}
    \includegraphics[width=8cm, clip, trim = 0cm 1.2cm 0cm 0cm]{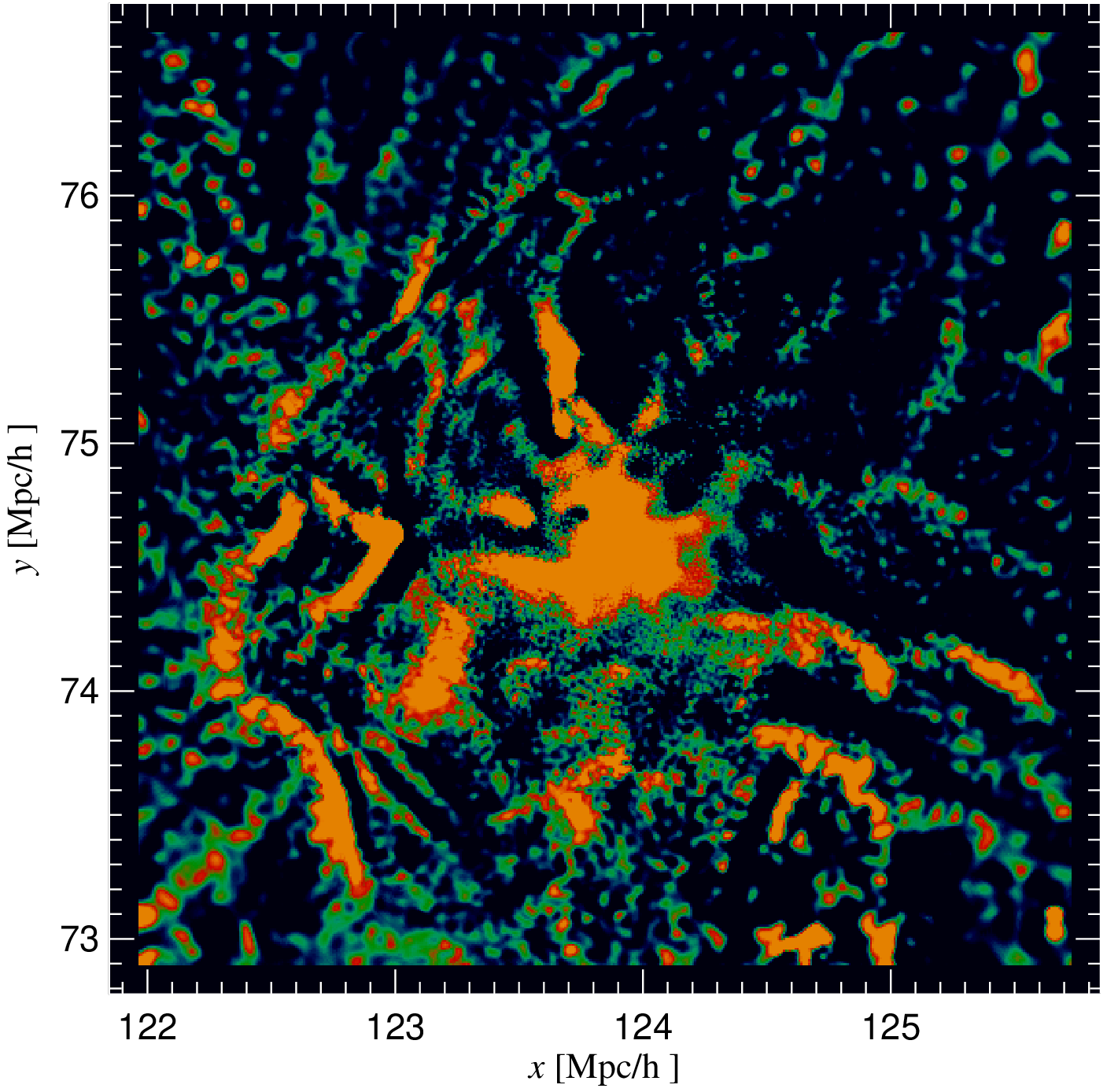}}
  \caption{\textbf{Pressure Discontinuities in SPH.} Projected pressure maps 
    in the SPH (top left) and SPHS (bottom left) $\times$ 256 runs within a 
    4 $h^{-1} {\rm Mpc}$ cube centred on the cluster at $z=0$. SPH produces
    sharp pressure discontinuities associated with orbiting substructures; 
    these are evident in the right panels, which are obtained by 
    unsharp masking (i.e. subtracting a smoothed version of the projected 
    pressure map to highlight residual structures).
  }
  \label{fig:ClusterCorePressure}
\end{figure*}

At high redshift, $z \sim 1$, there are already significant differences between the codes. 
Although SPH agrees qualitatively with SPHS and AMR at these early times, there is 
significantly more scatter between simulations of differing resolution. As the resolution 
is increased, the entropy core in SPH fluctuates significantly in amplitude by a factor up 
to $\sim 150$ (cf. Figure~\ref{fig:Redshift_Profiles}), suggesting 
non-convergent behaviour. This can be traced to the spurious surface tension reported 
originally in \citet{2007MNRAS.380..963A}. As detailed in \citet{2010MNRAS.405.1513R} and 
\citet{2012MNRAS.422.3037R}, this owes to multi-valued pressures at phase boundaries, such 
as arise when substructures containing lower entropy gas pass through the cluster core,
shocking and stirring the gas.
Since these drive pressure waves through the fluid, this propagates numerical errors away 
from regions of converging flow to the whole fluid domain (cf. Figure~\ref{fig:ClusterCorePressure}; 
pressure discontinuities are more pronounced in the SPH run). By contrast, in SPHS we introduce 
numerical 
dissipation when the flow is converging {\it designed to ensure single valued pressures} 
(and indeed to ensure all fluid quantities are single-valued). This keeps errors local, 
ensuring that they shift to smaller scales with increasing numerical resolution and, 
thereby, guaranteeing numerical convergence.

At low redshift $z=0$, the SPH results appear to converge on an ever lower central 
entropy. However, this illusion of convergence is actually driven by low entropy gas that 
artificially sinks to the cluster centre, protected by its numerical surface tension. This 
is masked at high redshift by on-going mergers that drive shocks and entropy generation in 
the gas. 
Note that dialling the entropy dissipation in SPHS down to zero, we find results that are 
similar to those from classic SPH (see figure \ref{fig:Entropy_Profiles_Diss}). This 
demonstrates that the differences between SPHS and AMR are driven largely by the numerical 
dissipation rather than the improved force accuracy in SPHS.

\subsection{Do resolved scales in non-radiative 
simulations of galaxy cluster formation care about the details of 
dissipation (physical or numerical) on unresolved scales?}\label{sec:dissres}

A key advantage of SPHS is that we can control the numerical dissipation, dialling it both 
up and down. This allows us to measure the impact of unresolved dissipative processes on 
resolved scales in the simulations. It has already been reported that numerical 
dissipation on small scales can affect the size and magnitude of a central entropy core 
\citep{2008MNRAS.387..427W}. However, there is a key difference between the dissipation 
added in \citet{2008MNRAS.387..427W} and that in SPHS. \citet{2008MNRAS.387..427W} build a 
simple sub-grid model for unresolved turbulence as a physical driver of dissipation at the 
resolution limit. Similar but more sophisticated attempts at the same have also be 
conducted by \citep{2009ApJ...707...40M}. Such sub-grid turbulence acts everywhere in the 
simulation and appears at first sight desirable as it seeks to capture unresolved physics. 
Indeed, \citet{2009ApJ...707...40M} report a higher normalisation for entropy cores in 
their simulations that include a sub-grid turbulence model, suggesting that such sub-grid 
turbulence may well impact on resolved scales in galaxy cluster simulations. However, 
\citet{2009ApJ...707...40M} do not perform any numerical convergence tests. Thus it is not 
clear whether the entropy core they report in either case -- with or without sub-grid 
turbulence -- is a numerically robust solution. Indeed, \citet{2008MNRAS.387..427W} show 
that the amplitude of the entropy core that they form, at fixed numerical dissipation 
parameter, decreases with increasing resolution\footnote{The effect is smaller if 
small-scale waves are omitted from the higher resolution simulation, but convergence is 
not convincingly shown. There are also some oddities. With a very large diffusion 
coefficient ($C$=10) at fixed resolution, they actually form a {\it lower} amplitude core than that 
formed with intermediate values (see their figure 12). Although such values for
the diffusion coefficient are unphysically large, this counter-intuitive behaviour 
may reflect the limitations of the simplified sub-grid turbulence model employed.}. 
This underscores the key problem with sub-grid turbulence models: there is no guarantee 
that they will produce a faithful convergence on the continuum Euler equations. By 
contrast, the dissipation in SPHS is {\it numerical}. It is required in order to ensure 
single-valued fluid quantities throughout the flow, but is otherwise kept to a minimum. 
The situation is similar in the {\tt RAMSES} code where minimal (and therefore 
unavoidable) numerical dissipation follows from the Riemann solver 
\citep[e.g.][]{1979JCoPh..32..101V}. In both cases, we expect a rigorous convergence on 
the continuum Euler equations with increasing resolution.

The works of \citet{2008MNRAS.387..427W} and \citet{2009ApJ...707...40M} leave a dangling 
question mark over whether or not it is useful -- or indeed essential -- to build 
physically motivated sub-grid turbulence models, or whether we can be satisfied with 
simply keeping numerical dissipation to a minimum and performing numerical convergence 
studies. We can address this point using SPHS by dialling up $\alpha_\mathrm{max}$ to 
large values and seeing how this impacts results on resolved scales. This is shown in 
figure \ref{fig:Entropy_Profiles_Diss}. Notice that the results for the entropy profile of 
the gas are in excellent agreement even for very large values of $\alpha_\mathrm{max} = 
5$. Visual inspection of the gas density profiles show that the $\alpha_\mathrm{max} = 5$ 
simulation is significantly more dissipative than the $\alpha_\mathrm{max} = 1$ default 
case. However, such dissipation shifts to smaller scales with increasing resolution and 
the equivalent comparison at $\times 32$ resolution shows even fewer differences: the 
results for SPHS converge independently of our choice of $\alpha_\mathrm{max}$ (cf. 
Figure~\ref{fig:SPHS_alphamax}). Furthermore -- despite the very different nature of the 
errors, error propagation, and numerical dissipation -- the SPHS simulations converge on a 
solution in remarkable accord with the AMR simulation (see figure 
\ref{fig:Redshift_Entropy_Profiles_SPH_vs_SPHS}).

Our results suggest that while numerical dissipation is necessary in any numerical method, 
so long as it is kept to a minimum its effect on non-radiative galaxy cluster simulations 
is benign. Furthermore, there appears to be no requirement to physically model sub-grid 
dissipation processes. Indeed, doing so may even be undesirable if it leads to a spurious 
transfer of information from unresolved to resolved scales. This could spoil convergence, 
preventing a correct solution of the Euler equations in the continuum limit.

\subsection{What is the role of gravitational shock heating as an entropy generation 
mechanism in galaxy clusters?}

Real galaxy clusters in the Universe are known to split into two types: CC and NCC (see 
\S\ref{sec:intro}). Armed with our results from SPH, SPHS and AMR we can now return to the 
question of the physical origin of this dichotomy. It is clear that in the absence of 
radiative cooling, entropy cores consistent with NCC clusters form, with the entropy 
generated from shocked gas during the cluster assembly process. It is likely, however, 
that real NCC clusters result from a more complex interplay between heating and cooling in 
the cluster core \citep{2008MNRAS.386.1309M}. While it is beyond the scope of this work to 
fully explain the observed dichotomy between NCC and CC clusters in nature, we have laid 
the foundations for such a study. Understanding the numerically well-defined problem of 
non-radiative galaxy clusters allows us to move with confidence to more physically 
realistic simulations that model also cooling, star formation and feedback from supernovae 
and active galactic nuclei. This will be the subject of forthcoming papers.

\section{Conclusions}
\label{sec:conclusions}

We have studied the entropy profile of the intracluster medium in a massive galaxy cluster 
forming in a non-radiative hydrodynamical cosmological resimulation using classic SPH, 
SPHS and AMR codes. In common with previous studies, we find that SPH entropy profiles 
decline with decreasing cluster-centric radius, whereas SPHS and AMR entropy profiles are 
in excellent agreement, plateauing to a well-defined value. Our key conclusions are as 
follows:

\begin{enumerate} 

\item The classic SPH result is incorrect, owing to a known artificial surface tension 
that appears at phase boundaries. At early times, the passage of massive substructures 
close to the cluster centre shock and stir gas, building up an entropy core.
At late times, the artificial surface tension causes low entropy gas -- that ought to mix with the higher entropy gas -- to sink 
artificially to the centre of the cluster.

\item Provided numerical dissipation occurs only at the resolution limit, and provided 
that it does not propagate errors to larger scales, we find that the effect of numerical 
dissipation is benign. There is no requirement to build `sub-grid' models of unresolved 
turbulence for galaxy cluster simulations.

\item Entropy cores in non-radiative simulations of galaxy clusters are physical, 
resulting from entropy generation in shocked gas during the cluster assembly process. This 
finally puts to rest the long-standing puzzle of cluster entropy cores in AMR simulations 
versus their apparent absence in classic SPH simulations. \end{enumerate}

\section*{Acknowledgements}
CP thanks Aaron Robotham for helpful conversations during the writing of this
paper.
The simulations presented in this paper were run on the iVEC EPIC and NCI VAYU
supercomputers. The research presented in this paper was undertaken as 
part of the Survey Simulation Pipeline (SSimPL; 
{\texttt{http://ssimpl-universe.tk/}). JIR acknowledges support 
from SNF grant PP00P2\_128540/1.

\vspace{0.1cm} \bsp

\bibliographystyle{mn2e}

\label{lastpage}

\end{document}